\documentclass[conference,10pt]{IEEEtran}

\usepackage{tcolorbox}
\usepackage{algorithm}
\usepackage[noend]{algpseudocode}
\usepackage{bm}
\usepackage{stackrel}
\usepackage{subfigure}
\usepackage{epsf}
\usepackage{framed}
\usepackage{amsmath,amssymb}
\usepackage{graphicx}
\usepackage{bbm}
\usepackage{color}
\usepackage{cite}
\usepackage{multirow,tabularx}
\usepackage{ifthen}
\usepackage{epstopdf}
\input{epstopdf.sty}
\usepackage{hyperref}
\usepackage{times}
\usepackage{caption} 
\usepackage{multirow}
\input{epsf.sty}

\makeatletter
\def\BState{\State\hskip-\ALG@thistlm}
\makeatother

\newcommand{\hide}[1]{\ifthenelse{\boolean{false}}{#1}{}}

\newtheorem{theorem}{{\bf Theorem}}

\newtheorem{lemma}{{\bf Lemma}}

\newtheorem{corollary}{{\bf Corollary}}

\newenvironment{definition}[1][Definition]{\begin{trivlist}
\item[\hskip \labelsep {\bfseries #1}]}{\end{trivlist}}

\newcommand{\qed}{\nobreak \ifvmode \relax \else
      \ifdim\lastskip<1.5em \hskip-\lastskip
      \hskip1.5em plus0em minus0.5em \fi \nobreak
      \vrule height0.75em width0.5em depth0.25em\fi}

\newcommand{\beq}{\begin{equation}}
\newcommand{\eeq}{\end{equation}}
\newcommand{\barr}{\begin{array}}
\newcommand{\earr}{\end{array}}

\newcommand{\benum}{\begin{enumerate}}
\newcommand{\eenum}{\end{enumerate}}

\newcommand{\bit}{\begin{itemize}}
\newcommand{\eit}{\end{itemize}}

\newcommand{\bc}{\begin{center}}
\newcommand{\ec}{\end{center}}

\newcommand{\bdes}{\begin{description}}
\newcommand{\edes}{\end{description}}

\newcommand{\bfig}{\begin{figure}}
\newcommand{\efig}{\end{figure}}

\newcommand{\bemq}{\begin{quote} \begin{em}}
\newcommand{\eemq}{\end{em} \end{quote}}

\newcommand{\bmp}{\begin{minipage}}
\newcommand{\emp}{\end{minipage}}

\newcommand{\supth}{^{{\mathrm{th}}}}

\newcommand{\bsp}{\begin{slide*}}
\newcommand{\esp}{\end{slide*}}
\newcommand{\bsl}{\begin{slide}}
\newcommand{\esl}{\end{slide}}

\newcommand{\blem}{\begin{lemma}}
\newcommand{\elem}{\end{lemma}}
\newcommand{\bthm}{\begin{theorem}}
\newcommand{\ethm}{\end{theorem}}

\IEEEoverridecommandlockouts

\begin{document}

\title{A Whittle Index Approach to Minimizing Functions of Age of Information}
\author{Vishrant Tripathi and Eytan Modiano\\
Laboratory for Information \& Decision Systems, MIT\thanks{This work was supported by NSF Grants AST-1547331, CNS-1713725, and CNS-1701964, and by Army Research Office (ARO) grant number W911NF-17-1-0508.}}

\IEEEaftertitletext{\vspace{-0.6\baselineskip}}
\maketitle
\begin{abstract}
We consider a setting where multiple active sources send  real-time updates over a single-hop wireless broadcast network to a monitoring station. Our goal is to design a scheduling policy that minimizes the time-average of general non-decreasing cost functions of Age of Information.
We use a Whittle index based approach to find low complexity scheduling policies that have good performance. We prove that for a system with two sources, having possibly different cost functions and reliable channels, the Whittle index policy is exactly optimal. We derive structural properties of an optimal policy, that suggest that the performance of the Whittle index policy may be close to optimal in general. These results might also be of independent interest in the study of restless multi-armed bandit problems with similar underlying structure. We further establish that minimizing monitoring error for linear time-invariant systems and symmetric Markov chains is equivalent to minimizing appropriately chosen monotone functions of Age of Information. Finally, we provide simulations comparing the Whittle index policy with optimal scheduling policies found using dynamic programming, which support our results.
\end{abstract}

\section{Introduction}

Many emerging applications depend on the timely delivery of status updates from a number of sources to a central monitor over a single-hop wireless network. Examples include sensor and actuator data for networked control systems, collecting information for IoT applications, mobility data in vehicular networks, and real-time surveillance and monitoring.

Age of Information (AoI) is a metric that captures timeliness of received information at a destination~\cite{kaul2012real,yin17_tit_update_or_wait}. Unlike packet delay, AoI measures the lag in obtaining information at a destination node, and is therefore suited for applications involving gathering or dissemination of time sensitive updates. Age of information, at a destination, is defined as the time that has elapsed since the last received information update was generated at the source. AoI, upon reception of a new update packet, drops to the time elapsed since generation of the packet, and grows linearly otherwise. Over the past few years, there has been a rapidly growing body of work on analyzing AoI for queuing systems \cite{kaul2012real,yin17_tit_update_or_wait,bedewy2019minimizing, huang2015optimizing, inoue2018general, kam2018age}, and using AoI as a metric for scheduling policies in networks \cite{kadota2016minimizing,kadota2018scheduling,kadota2018scheduling2,talak2018optimizing,tripathi2017age,jhun2018age,hsu2017scheduling, farazi2018age}.

The problem of minimizing age of information in single-hop networks was first considered in \cite{kadota2016minimizing} and \cite{kadota2018scheduling}. In these works, the authors considered a base station collecting time-sensitive information from a number of sources over a wireless broadcast network, where only one source can send an update at any given time. They looked at weighted linear combinations of AoI of all sources as the metric to be optimized. This prompted the design of low complexity scheduling policies that provably minimize weighted sum AoI at the base station, up to a constant multiplicative factor. These results crucially depend on the fact that for linear AoI, one can find a stationary randomized policy that is factor-2 optimal. As we will see later, this observation does not hold for general functions of AoI. In fact, stationary randomized policies can be arbitrarily worse than simple heuristic policies.

Scheduling problems with weighted linear combinations of age have also been considered with throughput constraints in \cite{kadota2018scheduling2} and with general interference constraints in 
\cite{talak2018optimizing}. 
AoI-based scheduling with stochastic arrivals was considered in \cite{hsu2017scheduling}, where a Whittle Index policy was shown to have good performance.

On the other hand, nonlinear cost functions of age were introduced as a natural extension to the AoI metric in \cite{yin17_tit_update_or_wait} for characterizing how the level of
dissatisfaction depends on data staleness in a more general manner. Nonlinear functions of age of information were also discussed in the context of queuing systems in \cite{kosta2017age} and \cite{kosta2018cost}. These papers develop the notion of value of information and use nonlinear cost of update delays, which correspond to nonlinear age cost functions. 

Nonlinear functions of age have also been discussed in the context of networked control systems in \cite{champati2019performance},\cite{klugelaoi} and \cite{ayan2019age}. In \cite{champati2019performance}, the authors discuss a real time networked control system and show that  the cost function is characterized as a non-decreasing, possibly nonlinear, function of AoI. In \cite{klugelaoi}, the authors formulated the state estimation problem for an LTI system, where the state of a discrete-time LTI system can be observed in any time-slot by paying a fixed transmission cost. The problem of minimizing the time-average of the sum of the estimation error and transmission cost reduces to minimizing a non-decreasing age-cost function for a single source with a fixed transmission cost. We explore this relationship more closely in Section \ref{sec:appl}, where we establish a similar equivalence for monitoring multiple LTI systems. 

\begin{figure}
\centering
\includegraphics[width=1.05\linewidth]{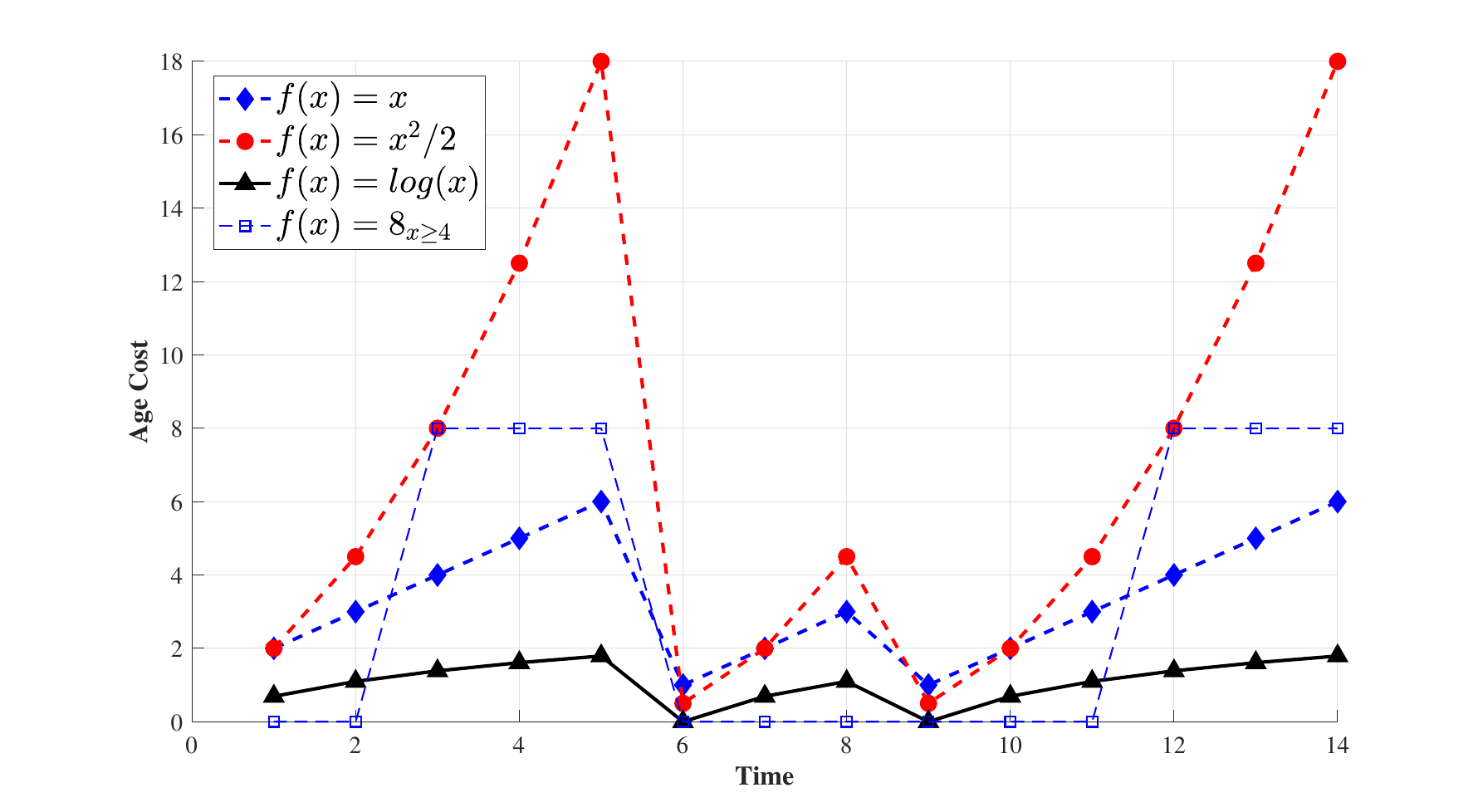}
\caption{Linear, quadratic, logarithmic and indicator cost functions for a sample age process. The linear process tracks the actual values of AoI.}
\label{figure:cost_ex}
\vspace{-3mm}
\end{figure}

In this work, we consider a setting similar to the one in \cite{kadota2016minimizing} and \cite{kadota2018scheduling}. We look at wireless broadcast network with $N$ sources generating real-time updates that need to be sent to a monitoring station. In any time-slot, only one source can attempt a transmission to the base station. Instead of weighted sum AoI, we are interested in minimizing the time-average of \textit{general non-decreasing cost functions} of AoI, summed over all sources. Examples of such functions include $f(x) = 2^{x}$, $f(x) = \log(x)$, $f(x) = \mathbbm{1}_{ \{x \geq 10 \} }$, etc. See Fig.\ref{figure:cost_ex} for examples. We develop a restless mutli-armed bandit formulation for the problem and use a Whittle Index based approach to find low complexity scheduling policies that have good performance.


Scheduling to minimize functions of age has also been considered in \cite{tripathi2017age} and \cite{jhun2018age}. In \cite{tripathi2017age}, the authors deals with minimizing symmetric functions of age of sources over multiple orthogonal unreliable channels and show that simple greedy schemes are asymptotically optimal. In \cite{jhun2018age}, the authors formulate the general functions of age problem with reliable channels and develop a high complexity algorithm that achieves minimum age. They also derive a key structural property of the optimal policy in this setting - the optimal policy is always periodic. However, their approach does not extend to the setting with unreliable channels. In this work, we consider unreliable channels and also build upon results from \cite{jhun2018age} and \cite{hsu2017scheduling} to derive stronger structural properties for optimal policies. These properties hint at why the performance of the heuristic Whittle index policy may be close to optimal. Moreover, it has been shown recently that the Whittle policy is indeed asymptotically optimal for \textit{linear} functions of AoI \cite{maatouk2020optimality}.

The remainder of the paper is organized as follows. In Section~\ref{sec:model}, we describe the general system model. In Section~\ref{sec:rmab}, we describe the equivalent restless multi-armed bandit formulation and discuss why we use the Whittle Index approach to solve the problem. In Section~\ref{sec:wrc}, we discuss the functions of age problem with reliable channels, develop the Whittle Index solution for this setting, and also prove key structural properties that an optimal policy must satisfy. In Section~\ref{sec:wurc}, we find the Whittle Index policy for the functions of age problem with unreliable channels. In Section~\ref{sec:sim}, we provide simulation results that verify our theoretical results. In Section~\ref{sec:appl}, we show that the problem of minimizing monitoring error for linear time-invariant systems when observing them over a wireless channel is equivalent to minimizing functions of AoI. We also show a similar result for monitoring symmetric Markov chains over a wireless channel. This shows the direct applicability of our Whittle framework to a large class of wireless monitoring problems.

A preliminary version of this paper appeared in the conference proceedings of Allerton 2019 \cite{tripathi2019whittle}. 

\section{Model}
\label{sec:model}
Consider a single-hop wireless network with $N$ active sources generating real-time status updates that need to be sent to a base station. We consider a slotted system in which each source takes a single time-slot to transmit an update to the base station. Due to interference, only one of the sources can transmit in any given time-slot.

For every source $i$, the age of information at the base station $A_i(t)$ measures the time elapsed since it received a fresh information update from the source. We assume active sources, i.e. in any time-slot, sources can generate fresh updates at will.  Let $s(t)$ be the source activated in time-slot $t$ and $u_i(t)$ be a Bernoulli random variable with parameter $p_i$ that denotes channel reliability between the $i^{th}$ source and the base station. Then, we have

\begin{figure}
\centering
\includegraphics[width=0.8\linewidth]{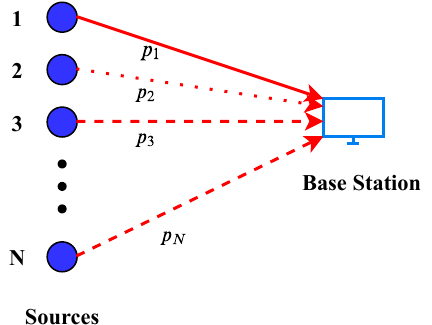}
\caption{$N$ sources transmitting updates to a base station over a wireless channel, with different reliabilities.}
\label{figure:cartoon}
\vspace{-5mm}
\end{figure}

\begin{equation}
A_i(t+1) =
 \begin{cases}
      A_i(t)+1, & \text{if }s(t)\neq i\text{ or }u_i(t) = 0, \\
      1, &  \text{if }s(t) = i \text{ and }u_i(t) = 1.
 \end{cases}
\end{equation}

In this work, we consider general cost functions of age as our metric of interest. For each source $i$, let $f_i(\cdot)$ denote a positive \textit{non-decreasing} cost function. 

Let $\pi$ be a scheduling scheme that decides which sources to schedule in every time-slot. The age process $A_i(t)$ depends on $\pi$ and the channel processes. Then, the expected average cost of age for source $i$ is given by

\begin{equation}
	\label{eq:main_prob}
C_i^{\text{ave}}(\pi) \triangleq \limsup_{T \rightarrow \infty} \frac{1}{T} \mathbb{E} \bigg[ \sum_{t = 1}^{T} f_i(A_i^{\pi}(t)) \bigg],
\end{equation}
where $A_i^{\pi}(t)$ is age process for the $i\supth$ source under policy $\pi$.

Our goal is find a schedule $\pi$ that minimizes the sum of average costs of age of sources, i.e. \eqref{eq:main_prob}. Let $\Pi$ denote the set of causal scheduling policies, then we want to solve the following optimization problem
\begin{equation}\label{eq:opt_def}
C^{\ast} = \min_{\pi \in \Pi} \sum_{i=1}^{N}C^{\text{ave}}_i(\pi),
\end{equation}
where $C^{\ast}$ is minimum average cost and $\pi^{\ast}$ is an optimal scheduling policy.

\section{Restless Multi-Armed Bandit Formulation}
\label{sec:rmab}

The restless multi-armed bandit (RMAB) is a classical resource allocation problem that has been studied in the optimization and operations research community for many decades. It involves $N$ Markov bandits, each of which evolves according to two different transition laws - one for when it is active, and one for when it is not. The scheduler can only activate one arm at any given time-slot, and a cost function maps the states of the arm to a corresponding cost in each time-slot. The goal of the scheduler is to design an arm activation policy that minimizes the long term time-average cost. The general solution strategy for such resource allocation problems is dynamic programming. However, that suffers from the curse of dimensionality and is not computationally feasible. Whittle, in his seminal work \cite{whittle1988restless}, showed that RMABs admit low complexity heuristic solutions called the Whittle Index if they satisfy a special property called indexability. In this section, we will show that scheduling to minimize such a metric can be reformulated as a RMAB.

Consider a restless multi-armed bandit problem with $N$ arms. The state space for every arm $i$ is the set of positive integers $\mathbb{Z}^{+}$. The state evolution of the arm depends on whether it is currently active or not. Let the state of arm $i$ at time $t$ be denoted by $A_i(t)$. If arm $i$ is active in time-slot $t$ then the state evolution is given by 

\begin{equation}
A_i(t+1) =
 \begin{cases}
      A_i(t)+1, & \text{w.p. }1-p_i \\
      1, &  \text{w.p. }p_i.
 \end{cases}
\end{equation}

If the arm is not active in time-slot $t$, then the state evolution is given by 
\begin{equation}
A_i(t+1) =
      A_i(t)+1.
\end{equation}

For every arm $i$, there is a cost function $f_i:\mathbb{Z}^{+}\rightarrow \mathbb{R}^{+}$ which maps the states of the arm to their associated costs. Thus, the cost of a state $\bm{x} \in \mathbb{Z}^{+^{N}}$ is given by $\sum_{i=1}^{N} f_i(x_i)$, where $\bm{x}$ is a vector of states (ages) and $x_i$ is the state (age) of the $i\supth$ source. Given that only one arm can be activated in any time-slot, the goal of the RMAB framework is to find a scheduling policy that minimizes the total time average cost of running this system.

This establishes the equivalence between the functions of age problem discussed earlier and a corresponding restless multi-armed bandit. Observe that the ``restless'' part of our construction cannot be dropped, since the states of the arms do not freeze when they are not active and there is no way to reformulate our problem as a simple (non-restless) multi-armed bandit problem. If that were the case, we could have found an optimal policy by solving for the Gittins index \cite{gittins1989multi}. However, finding optimal policies for restless bandits is much harder. The usual approach is to find the Whittle Index policy which provides good performance under certain conditions, namely \textit{indexability} of the RMAB problem.


In \cite{kadota2016minimizing} and \cite{kadota2018scheduling}, the authors develop three methods to solve the minimum age scheduling problem. First, they look at stationary randomized policies, where a source $i$ is scheduled at random with a fixed probability $p_i$. They find a stationary randomized policy that is factor-2 optimal for weighted sum AoI. However, this result does not hold for general functions: even the best stationary randomized policies in our setting can lead to an unbounded overall cost, despite there being very simple policies that have bounded cost. We demonstrate this with a simple example.

Consider two identical sources with cost functions given by $f(x) = 3^{x}$ and reliable channels, i.e $p_1 = p_2 = 1$. Any stationary randomized policy schedules at least one of the sources with probability less than or equal to $0.5$. For this source, the average cost is lower bounded by $\lim\limits_{T \rightarrow \infty}   \sum_{t=1}^{T}(3^t) \frac{0.5^{T}}{T}$ since with probability at least $0.5$, it does not get to transmit and its age increases by 1 in every time-slot. Observe that this lower bound goes to $\infty$ and hence the average cost also goes to $\infty$ for all stationary randomized policies. On the other hand, a simple round-robin scheme that schedules the two sources in alternating time-slots guarantees bounded cost for both sensors. Thus, stationary randomized policies can be infinitely worse than the optimal policy for the functions of age problem.

The second method developed for age-based scheduling in \cite{kadota2016minimizing,kadota2018scheduling} uses a Max-Weight approach. The authors design a quadratic Lyapunov function for the weighted sum of linear functions of AoI and find the max-weight policy - the policy that maximizes the amount of negative drift in the Lyapunov function in every time-slot. Performance guarantees for the max-weight policy crucially rely on the fact that there exists a stationary randomized policy that is factor-2 optimal for linear functions of age. Since this is not the case for general functions of age, we cannot develop similar performance bounds using a Max-Weight policy for the general functions of age problem.

This finally leaves us with the third method - using a Whittle Index based approach. In the following two sections, we use the RMAB formulation to establish indexability for the functions of age problem and derive a Whittle Index policy. We also show that for the case with 2 sources and reliable channels, the Whittle index policy is exactly optimal. This is a novel result since the optimality of Whittle Index policies is typically shown either only asymptotically, or in symmetric settings for finite systems. On the other hand, our optimality result holds for two asymmetric sources. 
\section{Reliable Channels}
\label{sec:wrc}

We first look at the problem with reliable channels between the sources and the base station. This leads to simpler analysis and a better understanding of the problem. Consider the setup described in Section I with channel reliability $u_i(t) = 1$, for all $i$ and $t$. In other words, the probability of success $p_i = 1, \forall i$.

In Section~\ref{sec:rmab}, we showed that the functions of age minimization problem is equivalent to a restless multi-armed bandit problem. Next, we use a Whittle Index based approach to try and solve the problem.

The first step in the Whittle Index approach is to formulate the \textit{decoupled problem}, where we consider a single arm in isolation with a fixed charge required to activate the arm.

\begin{framed}
\begin{definition}
\textit{Decoupled Problem}\\
Consider a single arm with the state space $\mathbb{Z}^{+}$ and an associated non-decreasing cost function $f:\mathbb{Z}^{+} \rightarrow \mathbb{R}^{+}$. Let the state of the arm be $A(t)$. Its evolution is given by
\begin{equation*}
A(t+1) =
 \begin{cases}
      A(t)+1, & \text{if not active at time t} \\
      1, &  \text{otherwise}.
 \end{cases}
\end{equation*}
There is a strictly positive activation charge $C$ to be paid in every time-slot that the arm is pulled.
\end{definition}
\end{framed}

Our goal is to find a scheduling policy that minimizes the time-average cost of running this system. Assuming that the cost function $f(\cdot)$ is non-negative and non-decreasing, we solve the decoupled problem using dynamic programming. The case when the activation charge is set to zero is trivial. The optimal policy is to always activate the arm. So, we consider $C$ to be strictly positive. The single source decoupled problem has also been solved in a slightly different setting in \cite{klugelaoi}.

\begin{framed}
\begin{theorem}
\label{thm:wrc}
The optimal policy for the decoupled problem is a stationary threshold policy. Let $H$ satisfy
\begin{equation}
\label{eq:wrc_con}
    f(H) \leq \frac{\sum_{j=1}^{H}f(j) + C}{H} \leq f(H+1).
\end{equation}
Then, the optimal policy is to activate the arm at time-slot $t$ if $A(t) \geq H$ and to let it rest otherwise. If no such $H$ exists, the optimal policy is to never activate the arm.
\end{theorem}
\end{framed}

\begin{IEEEproof}
See Appendix~\ref{pf:thm_wrc}.
\end{IEEEproof}

Theorem~\ref{thm:wrc} establishes that the optimal policy for the decoupled problem has a threshold structure. We now want to show that the \textit{indexability} property also holds for the decoupled problem. The indexability property states that as the activation charge $C$ increases from $0$ to $\infty$, the set of states for which it is optimal to activate the arm decreases monotonically from the entire set $\mathbb{Z}^{+}$ to the empty set $\{\phi\}$.

\begin{framed}
\begin{theorem}
\label{thm:rc_rel}
The \textit{indexability} property holds for the decoupled problem.
\end{theorem}
\end{framed}

\begin{IEEEproof}
See Appendix~\ref{pf:thm_ind_1}.
\end{IEEEproof}

The Whittle index approach states that if the decoupled problem satisfies the indexability property, we can formulate a heuristic index policy called the Whittle Index Policy that has good performance.

\begin{framed}
\begin{definition}
\textit{Whittle Index}\\
Consider the decoupled problem and denote by $W(h)$ the Whittle index in state $h$. Given indexability, $W(h)$ is the infimum charge $C$ that makes both decisions (activate, not activate) equally desirable in state $h$. The expression for $W(h)$ is given by
\begin{equation}
    W(h) = hf\big(h+1\big) - \sum_{j=1}^{h}f(j).
\end{equation}
\end{definition}
\end{framed}

Observe that using \eqref{eq:wrc_con}, $C = W(h)$ is the minimum value of the activation charge that makes both actions equally desirable in state $h$. This gives us the expression for the Whittle index. 

Let $W_i(x) :=  x f_i\big(x+1\big) - \sum_{j=1}^{x}f_i(j)$ represent the index function for the $i\supth$ decoupled problem. By the monotonicity of $f_i(\cdot)$, it is easy to see that the functions $W_i(\cdot)$ are also monotonically non-decreasing. This is because $W_i(h) - W_i(h-1) = h \big(f_i(h+1) - f_i(h)\big) \geq 0, \forall h$ since $f_i(\cdot)$ is non-decreasing. Using these functions, we define the Whittle Index Policy.
\begin{framed}
\begin{definition}
\textit{Whittle Index Policy}\\
Let $\pi^{W}(t)$ be the action taken by the Whittle Index Policy at time $t$. Then $\pi^{W}(t)$ is given by
\begin{equation}
\begin{split}
    \pi^{W}(t) &= \text{arg}\underset{1 \leq i \leq N}{\text{max}} \bigg\{ W_i \big(A_i(t)\big) \bigg\}\\
    &= \text{arg}\underset{1 \leq i \leq N}{\text{max}} \bigg\{ A_i(t)f_i\bigg(A_i(t)+1\bigg) - \sum_{j=1}^{A_i(t)}f_i(j) \bigg\}.
\end{split}
\end{equation}
\end{definition}
\end{framed}

Consider the case when the cost functions are weighted linear functions of AoI, i.e let $f_i(A_i(t)) = w_i A_i(t)$, with positive weights $w_i$. This is the setting considered in \cite{kadota2016minimizing} and \cite{kadota2018scheduling}. The Whittle Index for source $i$ is then given by $W_i(A_i(t)) = w_i (A_i^{2}(t)+A_i(t))/2$. This is the same as the Whittle index found in \cite{kadota2016minimizing}, where the authors showed that the Whittle policy is optimal for symmetric settings when all the weights are equal. We also establish that for $N=2$, the Whittle index policy is optimal even for asymmetric settings.

\begin{framed}
\begin{theorem}
\label{thm:2_opt}
For the functions of age problem with reliable channels and two sources, the Whittle index policy is exactly optimal.
\ethm
\end{framed}

\begin{IEEEproof}
See Appendix~\ref{pf:thm_2_opt}.
\end{IEEEproof}
This is an atypical result for restless multi-armed bandit problems which typically only have optimality results for symmetric or asymptotic settings. Our result is valid for finite ($N=2$) asymmetric settings. To the best of our knowledge, this is the first work to prove such a result for a restless multi-armed bandit problem. Next, we discuss some general properties that an optimal policy satisfies even for larger size systems. These properties help us establish the optimality of the Whittle index policy for $N=2$ and provide insight as to why the Whittle index policy has good performance in general.

\subsection{Properties of an Optimal Policy}
For the functions of age problem, a policy is stationary if it depends only on the current values of age. A cyclic policy is one that repeats a finite sequence of actions in a fixed order. We  define the space of policies that are stationary and periodic.
\begin{framed}
\begin{definition}
\textit{Stationary Cyclic Policies}\\
A stationary cyclic policy is a stationary policy that cycles through a finite subset of points in the state space, repeating a fixed sequence of actions in a particular order.
\end{definition}
\end{framed}

In \cite{jhun2018age}, the authors show that for reliable channels there exists an optimal policy that is  stationary, cyclic and can be found by solving the minimum average cost cycle problem over a large graph.

We look at this cyclic policy and analyze its properties. If there are multiple such cycles, we consider a cycle with the shortest length. We denote the length of the cycle by $T$ and age vectors on the cycle to be $\bm{x_1},\dots,\bm{x_T}$. Let the corresponding scheduling decisions be $d_1,\dots,d_T$. This implies that for state $\bm{x_k}$, taking action $d_k$ leads to the state $\bm{x_{k+1}}$, where the subscripts cycle back to $1,2,\dots$ after $T$. 

We establish an important structural property that such an optimal policy must satisfy, which we call the \textit{strong-switch-type} property. We call the policies that satisfy this property \textit{strong-switch-type} policies.

\begin{framed}
\begin{definition}
\textit{Strong-switch-type Policies}\\
Consider a stationary policy $\pi$ that maps every point in the state space $\mathbb{Z}^{{+}^{N}}$ to the set of arms $\{1,\dots,N\}$. We say that such a policy is strong-switch-type if
\begin{equation*}
    \pi(x_1,\dots,x_N) = i
\end{equation*}
implies
\begin{equation*}
    \pi(x_1',\dots,x_N') = i,
\end{equation*}
for all $\bm{x}$ and $\bm{x'}$ such that $x_i' \geq x_i$ and $x_j' \leq x_j, \forall j \neq i$.
\end{definition}
\end{framed}

In words, the strong-switch-type property implies that if a policy decides to activate arm $i$ for a state vector $\bm{x}$, then for a state vector $\bm{x'}$ with a higher age for the $i\supth$ source and lower ages for all the other sources, it still decides to activate source $i$. Note that our definition of strong-switch-type policies is a stronger version of the switch-type policies introduced in \cite{hsu2017scheduling}.

\begin{framed}
\begin{theorem}
\label{thm:sw_cyc}
For the functions of age problem with reliable channels, all state-action pairs that are a part of the shortest length optimal cyclic policy must satisfy the strong-switch-type property.
\end{theorem}
\end{framed}
\begin{IEEEproof}
See Appendix~\ref{pf:sw_cyc}.
\end{IEEEproof}

We can prove this result for general values of $N$. However, to extend the strong-switch-type property over the entire state-space, we consider systems with up to three sources. 
\begin{framed}
\begin{theorem}
\label{thm:sw_opt}
There exists an optimal stationary policy for the functions of age problem with reliable channels and up to three sources that has the strong-switch-type property \textit{over the entire state-space}.
\end{theorem}
\end{framed}
\begin{IEEEproof}
We have already established that points on the minimum average cost cycle satisfy the strong-switch-type property. In Appendix~\ref{pf:sw_opt}, we extend this policy over the entire state space while maintaining the strong-switch property to obtain a well defined stationary policy.
\end{IEEEproof}
While we prove this result for up to three source and reliable channels, we believe that the strong-switch-type property is a natural property that an optimal policy must have in general, due to monotonicity of cost functions.

We now define the space of policies that can be found as a result of the Whittle Index based approach. 
 
 \begin{framed}
\begin{definition}
\textit{Index Policies}\\
Consider a stationary policy $\pi$ that maps every point in the state space $\mathbb{Z}^{{+}^{N}}$ to the set of arms $\{1,\dots,N\}$. We say that such a policy is an index policy if
\begin{equation*}
    \pi(x_1,\dots,x_N) = \text{arg}\underset{1 \leq i \leq N}{\text{max}} \bigg\{ F_i(x_i) \bigg\}
\end{equation*}
for all $\bm{x}$, where $F_i:\mathbb{Z}^{+}\rightarrow \mathbb{R}$ are monotonically non-decreasing functions for all $i$.
\end{definition}
\end{framed}
 Observe that if $F_i$ are the same as $W_i$ in the above definition, then we get back the Whittle Index Policy. Also, note that an index policy always satisfies the strong-switch-type property by definition. This is because the index functions $F_i(\cdot)$ are monotonically non-decreasing. We now show that index policies are in fact the same as strong-switch-type policies.
  
\begin{framed}
\begin{theorem}
\label{thm:eq_pol}
For the functions of age problem, every policy that is strong-switch-type is also an index policy.
\end{theorem}
\end{framed}

\begin{IEEEproof}
The proof is based on induction on the number of sources. We assume that every strong-switch-type policy can be represented as an index policy for systems with $N$ sources. Using this fact, we show that strong-switch-type policies can also be represented as index policies for systems with $N+1$ sources. We also show that the two types of policies are equivalent for the single source decoupled problem, thus completing the proof. The details are in Appendix~\ref{pf:thm_eq_pol}.
\end{IEEEproof}
An important point to notice is that while we use the reliability of channels in the proof of Theorem~\ref{thm:sw_opt}, we do not use any such condition for the proof of Theorem~\ref{thm:eq_pol}. Thus, strong-switch-type policies are equivalent to index policies regardless of channel connectivity.

Theorems \ref{thm:sw_opt} and \ref{thm:eq_pol} together imply the following corollary.
\begin{framed}
\begin{corollary}
\label{corr_opt}
For the functions of age problem with reliable channels and up to three source, there exists a stationary optimal policy that is an index policy.
\end{corollary} 
\end{framed}
In other words, there exists an optimal policy that looks like the Whittle Index policy in that the arm to be activated has the maximum value among \textit{monotone index functions} that take as arguments only the states of individual arms. This hints at why the performance of Whittle Index policies may be close to optimal.

Observe that the Whittle Index policy would be optimal in general if we could show that it achieves a cost that is the minimum cost among the space of index policies and that the strong-switch-type property holds for some optimal policy. We show that this is indeed the case for $N=2$. However, we  later provide an example that shows that the Whittle policy is not optimal, but only close to optimal, for $N=4$. 

We leave the question of whether the Whittle index policy is at most a constant factor away from optimal in general to future work. We believe that the structural properties introduced here provide a recipe to proving constant factor optimality of the Whittle index policy, even for general bandit problems with similar underlying structure.

\section{Unreliable Channels}
\label{sec:wurc}
We now consider independent Bernoulli channels between every source and the base station, with probability of success $p_i$ for source $i$. We derive a Whittle index in this setting and establish indexability of the RMAB problem by enforcing a bounded cost condition on the functions $f_i(\cdot)$.

An important fact to notice is that monotonicity in itself is not sufficient to ensure that the system has finite average cost even for $N=1$, in the case of unreliable channels. Consider a single source case where $f(a) = 3^{a}$ and the probability of success $p = 0.5$. If the source attempts a transmission in every time-slot, the expected average cost satisfies
\begin{equation}
 \limsup_{T \rightarrow \infty}   \sum_{t=1}^{T}(3^t) \frac{0.5^{T}}{T} \leq  \limsup_{T \rightarrow \infty} \frac{1}{T} \mathbb{E} \bigg[ \sum_{t = 1}^{T} 3^{A(t)} \bigg],
\end{equation}
since with probability $0.5$, the transmission fails and age increases by 1 in every time-slot. However, observe that the summation on the left goes to infinity and thus the expected average cost goes to infinity. This happens despite the source attempting a transmission in every time-slot. To prevent such a situation from happening we enforce the following \textit{bounded cost} condition on the age cost functions $f_i$ in addition to monotonicity 
\begin{equation}
    \sum_{h=1}^{\infty} f_i(h) (1-p_i)^{h} < \infty.
\end{equation}
It can be shown that this condition ensures that the single source case has bounded cost. We define the decoupled problem in this case as follows:

\begin{framed}
\begin{definition}
\textit{Decoupled Problem}\\
Consider a single arm with the state space $\mathbb{Z}^{+},$ probability of success $p$ and an associated non-decreasing cost function $f:\mathbb{Z}^{+} \rightarrow \mathbb{R}^{+}$ that satisfies the \textit{bounded cost} condition. Let the state of the arm be $A(t)$. If the arm is active at time $t$, its evolution is given by
\begin{equation*}
A(t+1) =
 \begin{cases}
      A(t)+1, & \text{w.p. }1-p \\
      1, &  \text{w.p. }p.
 \end{cases}
\end{equation*}
If the arm is not active in time-slot $t$, then the state evolution is given by 
\begin{equation*}
A(t+1) =
      A(t)+1.
\end{equation*}
There is a strictly positive activation charge $C$ to be paid in every time-slot that the arm is pulled.
\end{definition}
\end{framed}

As before, our goal is to find a scheduling policy that minimizes the time-average cost of running this system.

\begin{framed}
\begin{theorem}
\label{thm:wurc}
The optimal policy for the decoupled problem is a stationary threshold policy. Let $H$ satisfy
\begin{equation}
\label{eq:urc_con}
\begin{split}
   p^2 (H-1) &\bigg(\sum_{k=H}^{\infty} f(k)(1-p)^{k-H}\bigg) - p \bigg(\sum_{j=1}^{H-1}f(j)\bigg) \\ \leq &~C\\ \leq p^2 H \bigg(&\sum_{k=H+1}^{\infty} f(k)(1-p)^{k-H-1}\bigg) - p \bigg(\sum_{j=1}^{H}f(j)\bigg)
\end{split}
\end{equation}
Then, the optimal policy is to activate the arm at time-slot $t$ if $A(t) \geq H$ and to let it rest otherwise. If no such $H$ exists, the optimal policy is to never activate the arm.
\end{theorem}
\end{framed}

\begin{IEEEproof}
See Appendix~\ref{pf:thm_wurc}.
\end{IEEEproof}

Observe that taking the limit as $p \rightarrow 1$ in Theorem~\ref{thm:wurc}, we get back the threshold policy for reliable channels derived in Theorem~\ref{thm:wrc}. We now establish indexability and derive the functional form of the Whittle Index.

\begin{framed}
\begin{theorem}
\label{thm:urc_rel}
The \textit{indexability} property holds for the decoupled problem. Denote by $W(h)$ the Whittle index in state $h$. Given indexability, $W(h)$ is the infimum charge $C$ that makes both decisions (activate, not activate) equally desirable in state $h$. The expression for $W(h)$ is given by
\begin{equation}
    W(h) = p^2 h \big(\sum_{k=1}^{\infty}f(k+h)(1-p)^{k-1}\big) - p\big(\sum_{j=1}^{h}f(j)\big).
\end{equation}
\end{theorem}
\end{framed}

\begin{IEEEproof}
See Appendix~\ref{pf:thm_urc_rel}.
\end{IEEEproof}

Again, observe that taking the limit as $p\rightarrow1$, we get back the Whittle Index derived in Section~\ref{sec:wrc}. Further, if we assume that the cost functions are weighted linear functions of AoI, i.e. $f_i(A_i(t)) = w_i A_i(t)$ where all the weights are positive, then the index functions for the Whittle policy are given by $W_i(A_i(t)) = w_i p_i A_i(t) (A_i(t) + \frac{1 + (1-p_i)}{1-(1-p_i)})/2$. This corresponds to the index policy developed in \cite{kadota2016minimizing}, where the authors showed that for symmetric settings when all the weights and channels probabilities are equal, the Whittle index policy is optimal.

\section{Simulations}
\label{sec:sim}
First, we compare the optimal policy, found using dynamic programming, with the Whittle index policy for two sources. We consider six different settings in total -  3 sets of functions, each with reliable and unreliable channels. 

For settings $A_1$ and $A_2$, the cost functions are chosen to be $f_1(x) = 13x$ and $f_2(x) = x^2$. In $A_1$, we consider reliable channels, i.e. $p_1 = p_2 = 1$. In $A_2$, we consider unreliable channels, specifically $p_1 = 0.9$ and $p_2 = 0.5$. For settings $B_1$ and $B_2$, the cost functions are chosen to be $f_1(x) = x^2$ and $f_2(x) = 3^x$. In $B_1$, we consider reliable channels, i.e. $p_1 = p_2 = 1$. In $B_2$, we consider unreliable channels, specifically $p_1 = 0.65$ and $p_2 = 0.8$. For settings $C_1$ and $C_2$, the cost functions are chosen to be $f_1(x) = x^3/2$ and $f_2(x) = 10\log(x)$. In $C_1$, we consider reliable channels, i.e. $p_1 = p_2 = 1$. In $C_2$, we consider unreliable channels, specifically $p_1 = 0.55$ and $p_2 = 0.75$. Simulation results are presented in Table~\ref{table:1}.

\begin{table}[h!]
\centering
\begin{tabular}{|c|c|c|} 
 \hline
 Setting & Optimal Cost & Whittle Index Cost \\ [0.5ex] 
 \hline\hline
 $A_1$ (reliable) & 21.95 & 21.95 \\ 
 $A_2$ (unreliable) & 36.12 & 36.28 \\ \hline
 $B_1$ (reliable) & 8.48 & 8.48 \\
 $B_2$ (unreliable) & 23.16 & 23.37 \\ \hline
 $C_1$ (reliable) & 5.69 & 5.69 \\
 $C_2$ (unreliable) & 21.54 & 21.54  \\[0.5ex] 
 \hline
\end{tabular}
\caption{Cost of the Whittle index policy and the optimal dynamic programming policy for 2 sources.}
\label{table:1}
\end{table}

We find the optimal cost for each setting using finite horizon dynamic programming over a horizon of 500 time-slots. For reliable channels, we find the cost of the Whittle index policy by simply implementing it once over 500 time-slots. For unreliable channels, we estimate the expected Whittle index cost by averaging the performance of the Whittle index policy over 500 independent runs. 

Observe that the Whittle index policy is exactly optimal when the channels are reliable, as expected from our theoretical results. The expected cost for the Whittle index policy is very close to the optimal cost for unreliable channels as well. Also, for the same set of functions, having unreliable channels increases the cost compared to reliable channels, as expected.

Next, we compare the optimal policy with the Whittle index policy for more than two sources.  Simulation results are presented in Table~\ref{table:2}.

For settings $D_1$ and $D_2$, we consider 3 sources. The cost functions are chosen to be $f_1(x) = x^2$, $f_2(x) = 3^x$ and $f_3(x) = x^4$. In $D_1$, we consider reliable channels, i.e. $p_1 = p_2 = p_3 = 1$. In $D_2$, we consider unreliable channels, specifically $p_1 = 0.66$, $p_2 = 0.8$ and $p_3 = 0.75$. 

For settings $E_1$ and $E_2$, we consider 4 sources. The cost functions are chosen to be $f_1(x) = x^3$, $f_2(x) = 2^x$, $f_3(x) = 15x$ and $f_4(x) = x^2$. In $E_1$, we consider reliable channels, i.e. $p_1 = p_2 = p_3 = 1$. In $E_2$, we consider unreliable channels, specifically $p_1 = 0.7$, $p_2 = 0.9$, $p_3 = 0.67$ and $p_4 = 0.8$.

\begin{table}[h!]
\centering
\begin{tabular}{|c|c|c|p{1.5cm}|} 
 \hline
 No. of Sources & Setting & Optimal Cost & Whittle Index Cost \\ 
 \hline\hline
 \multirow{2}{*}{3} &  $D_1$ (reliable) & 44.23 & 44.23 \\ 
  & $D_2$ (unreliable) & 161.19 & 161.39 \\ \hline
 \multirow{2}{*}{4} &$E_1$ (reliable) & 73.36 & 73.36 \\
  & $E_2$ (unreliable) & 129.02 & 130.94 \\ \hline
 \multirow{2}{*}{4} &$F_1$ (reliable) & 87.66 & 88.27 \\
  & $F_2$ (unreliable) & 158.35 & 159.81  \\[0.5ex] 
 \hline
\end{tabular}
\caption{Cost of the Whittle index policy and the optimal policy for more than 2 sources.}
\label{table:2}
\end{table}

For settings $F_1$ and $F_2$, we consider 4 sources. The cost functions are chosen to be $f_1(x) = x^3$, $f_2(x) = e^x$, $f_3(x) = 15x$ and $f_4(x) = x^2$. In $F_1$, we consider reliable channels, i.e. $p_1 = p_2 = p_3 = 1$. In $F_2$, we consider unreliable channels, specifically $p_1 = 0.8$, $p_2 = 0.85$, $p_3 = 0.75$ and $p_4 = 0.66$.

We observe that the cost of the Whittle index policy is the same as that obtained using dynamic programming for settings $D_1$ and $E_1$. However, for setting $F_1$, we observe a small gap in performance between the two policies, thus giving us an example that shows that the \textit{Whittle index policy need not be optimal}, in general. We also verify that the optimal policy found using dynamic programming follows a cyclic pattern that satisfies the strong-switch-type property and is distinct from the Whittle index policy. This is also in line with our discussion on structural properties. 

We note that computing the optimal policy using dynamic programming becomes progressively harder in terms of space and time complexity for larger values of $N$, as the state-space to be considered grows exponentially with $N$. The Whittle index policy, on the other hand, is very easy to compute and implement with only a linear increase in space and time complexity with the number of sources. Also, as is evident from simulations, the performance of the Whittle policy is close to optimal in every setting considered, thus making it a very good low complexity heuristic.

\section{Applications}
\label{sec:appl}
In this section, we will apply the framework we have developed to two problems in remote monitoring and control to show that optimizing general functions of AoI arise naturally in many practical settings.

\subsection{Monitoring LTI systems}
First, we consider the remote monitoring of linear time-invariant (LTI) systems over a wireless channel. Suppose that there are $N$ such systems, where the $i$th system evolves over time as follows
\begin{equation}
	\label{eq:lti_evolution}
 {x_i(t+1)} = G_i {x_i(t)} + w_i(t),
\end{equation}
where $x_i(t) \in \mathbb{R}^{d_i}, G_i \in \mathbb{R}^{d_i \times d_i}$ is the system matrix and $w_i(t) \sim \mathcal{N}(0,\Sigma_i)$ is multi-variate zero-mean Gaussian noise, i.i.d. across time. We further assume that the noise increments $w_i(t)$ are independent across sources, so their evolution is decoupled.

Suppose that a central agent wants to monitor the state of each of the $N$ systems with as little monitoring error as possible. However, due to wireless interference constraints, it can only observe the state of one system at any given time-slot. How should the agent design a wireless scheduling policy that minimizes expected monitoring error?

Let $\hat{x}_i(t)$ represent the maximum likelihood estimate of the state of the $i$th system at the monitor at any given time-slot $t$, given past observations. We define monitoring error for the $i$th system as 
\begin{equation}
	\label{eq:error_def}
	e_i(t) \triangleq \mathbb{E} \bigg[ \big|\big|x_i(t) -  \hat{x}_i(t)\big|\big|^2_2  \bigg].
\end{equation}

The following theorem relates the expected monitoring error of the $i$th system to its AoI. Specifically, we compute the expected error if the $i$th system has not been observed for the last $\Delta$ time-slots.
\begin{framed}
	\begin{theorem}
		\label{thm:lti_fAoI}
		Suppose that the $i$th system evolves according to \eqref{eq:lti_evolution}. Further suppose that the monitor last observed the state of the $i$th system at time $t=\tau$.  Then, the expected monitoring error for the $i$th system at time $t=\tau+\Delta$ is given by
		\begin{equation}
			\label{eq:monitoring_functions}
			\begin{aligned}
				e_i(\tau+\Delta) &= \mathbb{E} \bigg[ \big|\big|x_i(\tau+\Delta) -  \hat{x}_i(\tau+\Delta)\big|\big|^2_2  \bigg]  \\&= \sum_{k=0}^{\Delta-1} Tr\big((G_i^k)^T (G_i^k) \Sigma_i\big) \triangleq f_i(\Delta). 
			\end{aligned}
		\end{equation}
	\end{theorem}
\end{framed}
\begin{IEEEproof}
	See Appendix~\ref{pf:lti_fAoI}.
\end{IEEEproof}

Using this observation, we can establish an equivalence between minimizing monitoring error and minimizing functions of AoI.  To find the scheduling policy $\pi$ that minimizes expected time-average monitoring error, we need to solve the following optimization problem
	\begin{equation}
		\begin{aligned}
			\min_{\pi \in \Pi} \limsup_{T \rightarrow \infty} \frac{1}{T}  \mathbb{E}\bigg[ \sum_{t = 1}^{T} \sum_{i=1}^{N} e_i(t) \bigg],
		\end{aligned}
	\end{equation}
where $e_i(t)$ is defined as in \eqref{eq:error_def}. This optimization problem is equivalent to solving the following functions of AoI problem
\begin{equation}
	\begin{aligned}
		\min_{\pi \in \Pi} \limsup_{T \rightarrow \infty} \frac{1}{T}  \bigg[ \sum_{t = 1}^{T} \sum_{i=1}^{N} f_i(A_i(t)) \bigg],
	\end{aligned}
\end{equation}
where $A_i(t)$ is the AoI of the $i$th system and the functions $f_i(\cdot)$ are as defined in \eqref{eq:monitoring_functions}.

We also show in Appendix~\ref{pf:lti_fAoI} that the functions $f_i(\cdot)$ are monotonically increasing, so we can indeed apply our Whittle index approach to solve this problem. The rate at which the functions $f_i(\cdot)$ increase depends on the eigenvalues of the system matrices $G_i$. If the largest eigenvalue of $G_i$ lies inside (outside) the unit circle, then $f_i(\cdot)$ increases slower (faster) than a linear function. If the largest eigenvalue of $G_i$ lies on the unit circle, then $f_i(\cdot)$ increases linearly.

\subsection{Monitoring Markov Chains}
\begin{figure}
	\centering
	\includegraphics[width=0.6\linewidth]{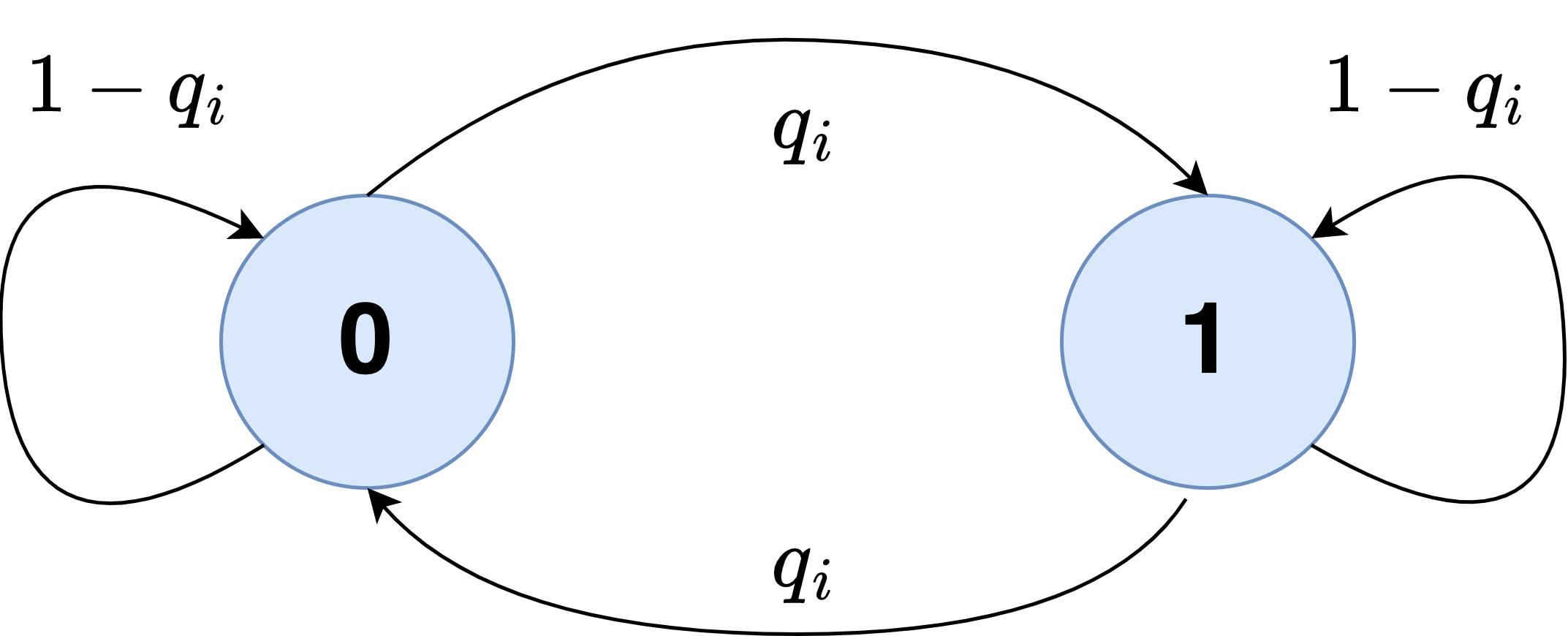}
	\caption{Symmetric two-state Markov chain, representing the state of the $i$th process.}
	\label{fig:markov}
	\vspace{-5mm}
\end{figure}

Consider $N$ symmetric two state Markov chains of the form drawn in Fig.~\ref{fig:markov} running in discrete-time. As for the previous example, we assume that only one system out of the $N$ can be observed in any given time-slot. We denote the distribution of the $i$th Markov chain at time $t$ by $x_i(t)$, where $x_i(t) = [1~~0]$ if the Markov chain is in state $0$ and $x_i(t) = [0~~1]$ if the Markov chain is in state $1$.

We assume that the base station knows the transition probability $q_i$ and the transition matrix $$Q_i = \begin{bmatrix}
	1-q_i & q_i \\
	q_i & 1-q_i
\end{bmatrix}$$ associated with the $i$th Markov chain and uses this to maintain the estimated distribution of the $i$th chain, based on the most recent observation. Suppose that the base station knew that the $i$th Markov chain had the distribution $x_i(\tau)$ at time $\tau$. Using the transition matrix $Q_i$ for the $i$th chain, the base station can compute the distribution of the Markov chain at time $\tau+\Delta$ given the information at time $\tau$. We denote this estimated distribution of the actual state by $\hat{x}_i(\tau+\Delta)$ and it is given by 
\begin{equation}
	\hat{x}_i(\tau+\Delta) = x_i(\tau) \begin{bmatrix}
		1-q_i & q_i\\
		q_i & 1-q_i
	\end{bmatrix}^\Delta = x_i(\tau) Q^\Delta_i.
\end{equation} 

We are interested in minimizing the monitoring error, defined as a notion of distance between the estimated distribution and the actual state of the Markov chain. We define error for the $i$th system as follows - 
\begin{equation}
	e_i(t) = \mathbb{E}\bigg[ D\big(x_i(t) || \hat{x}_i(t)\big)  \bigg],
\end{equation}
where $D$ is a notion of divergence between the two probability distributions. In this work, we will discuss our results for Kullback-Liebler (KL) divergence and total variation (TV) distance, however, the general ideas should work for other divergences as well. The KL divergence for discrete distributions is defined as 
$$D_{KL}(P || Q) = - \sum_{x \in \mathcal {X}} P(x) \log(\frac{P(x)}{Q(x)}).$$
The total variation (TV) distance for discrete distributions is defined as 
$$D_{TV}(P || Q) = \sum_{x \in \mathcal {X}} \frac{1}{2}\big|P(x) - Q(x)\big|.$$

The following theorem relates the expected monitoring error of the $i$th system to its AoI. Specifically, we compute the expected error if the $i$th system has not been observed for the last $\Delta$ time-slots.
\begin{framed}
	\begin{theorem}
		\label{thm:markov_fAoI}
		Suppose that the $i$th system evolves according to Markov chain in Fig.~\ref{fig:markov}. Further suppose that the monitor last observed the state of the system at time $t=\tau$.  Then, the expected monitoring error for the $i$th system at time $t=\tau+\Delta$ is given by
		\begin{equation}
			\label{eq:markov_functions}
			\begin{aligned}
				e_i(\tau&+\Delta) = \mathbb{E} \bigg[ D\big(x_i(\tau+\Delta) || \hat{x}_i(\tau+\Delta)\big)  \bigg]  \\&= \begin{cases}
					H\big([Q_i^\Delta]_{00}\big), &\text{ if } D \text{ is KL Divergence,}\\
					2[Q^\Delta_i]_{00}\big(1- [Q^\Delta_i]_{00} \big), &\text{ if } D \text{ is TV Distance}.
				\end{cases} \\&= f_i(\Delta).
			\end{aligned}
		\end{equation}
	Here $[Q_i^\Delta]_{00}$ is the top diagonal element of the transition matrix raised to the power $\Delta$, i.e. $Q^\Delta_i$ and $H(q)\triangleq -q \log(q) -(1-q)\log(1-q)$ is the binary entropy function.
	\end{theorem}
\end{framed}
\begin{IEEEproof}
	See Appendix~\ref{pf:markov_fAoI}.
\end{IEEEproof}
Using the result above, it is straightforward to establish an equivalence between minimizing monitoring error for Markov chains and minimizing functions of AoI. As for the case with the LTI systems, we further show in Appendix~\ref{pf:markov_fAoI} that the functions $f_i(\cdot)$ are monotonically increasing, so we can indeed apply our Whittle index approach to solve this problem.

An interesting observation for the KL divergence case is that the monitoring error cost ends up being the entropy of the estimated distribution of the Markov chain. This can be interpreted as the amount of uncertainty that the base station has about the Markov chain, which increases with the number of time-slots that the chain remains unobserved. We use this Markov model and our Whittle framework to solve a robotics problem involving time-varying multi-agent occupancy grid mapping in \cite{tripathi2021computation}.


\section{Conclusion}

In this work, we presented the problem of minimizing functions of age of information over a wireless broadcast network. We used a restless multi-armed bandit approach to establish indexability of the problem and found the Whittle index policy. For the case with two sources and reliable channels, we were able to show that the Whittle index policy is optimal. We also established structural properties of an optimal policy, for the case with reliable channels. These properties hint at why the performance of the Whittle index policy is close to optimal in general. 

A possible direction of future work is to try and prove constant factor optimality of the Whittle index policy in general, using the structural properties developed in this work. Other interesting extensions could be to consider sources with stochastic arrivals instead of active sources and handling unknown, possibly time-varying functions of Age of Information.

\bibliographystyle{ieeetr}
\bibliography{bibliography}

\appendix

\subsection{Proof of Theorem \ref{thm:wrc}}
\label{pf:thm_wrc}
Consider the decoupled problem described in Section~\ref{sec:wrc}. Let $u(t)$ be an indicator variable that denotes whether the arm is pulled or not at time $t$. Under a scheduling policy $\pi$ that specifies the value of $u(t)$ for all instants of time, the average cost is given by

\begin{equation}
    \lim_{T \rightarrow \infty}\frac{1}{T}\sum_{t=1}^T \bigg[f\big(A^{\pi}(t)\big) + C u^{\pi}(t) \bigg].
\end{equation}
 We want to find a policy that minimizes this cost over the space of all policies. Let $S:\mathbb{Z}^{+}\rightarrow\mathbb{R}$ denote the differential cost-to-go function for this problem, let $u:\mathbb{Z}^{+}\rightarrow \{1,0 \}$ be the stationary optimal policy and let $\lambda$ denote the optimal cost. Then, the Bellman equations are given by
 \begin{equation}
     S(h) = f(h) + \underset{u(h) \in \{1,0\}}{\text{min}} \{C,S(h+1)\} - \lambda, \forall h \in \mathbb{Z}^{+}.
 \end{equation}
Without loss of generality we set $S(1) = 0$. Assume that the optimal policy has a threshold structure, i.e. there exists $H$ such that it is optimal to pull the arm $(u(h)=1)$ for all states $h \geq H$ and let it rest otherwise $(u(h)=0)$. If this the case, then the Bellman equations reduce to
\begin{equation}
\label{eq:bell1}
    \begin{split}
    S(h) & = f(h) + C - \lambda, \forall h \geq H.
    \end{split}
\end{equation}
Using the monotonicity of $f(\cdot)$, we conclude that $S(h+1)\geq S(h), \forall h \geq H$. We will use this fact later. For the state $H-1$, we get
\begin{equation}
\begin{split}
S(H-1)& = f\big(H-1\big) - \lambda + S(H)\\
      & = f\big(H-1\big) - \lambda + f(H) - \lambda + C.
\end{split}
\end{equation}
Repeating this $k$ times, we get
\begin{equation}
\label{eq:below_th}
\begin{split}
S(H-k)& = \sum_{j=0}^{k}f\big(H-j\big) - (k+1) \lambda + C, 
\end{split}
\end{equation}
for all $k$ in $\{1,\dots,H-1\}$. Observe that since we set $S(1) = 0$, we get \begin{equation}
\label{eq:lam}
    \lambda = \frac{\sum_{j=1}^{H}f(j) + C}{H},
\end{equation}
by putting $k = H-1$ in \eqref{eq:below_th}. Now assume that $H$ further satisfies the relation given in Theorem~\ref{thm:wrc}, i.e.
\begin{equation}
\label{eq:wc}
    f(H) \leq \frac{\sum_{j=1}^{H}f(j) + C}{H} \leq f(H+1).
\end{equation}
Using \eqref{eq:lam}, we can simplify \eqref{eq:wc} as 
\begin{equation}
\label{eq:lamf}
    f(H) \leq \lambda \leq f(H+1).
\end{equation}
Adding $C-\lambda$ to every term above, we get 
\begin{equation}
\begin{split}
    & f(H) + C - \lambda \leq C \leq f(H+1) + C - \lambda\\
    \implies & S(H) \leq C \leq S(H+1). 
\end{split}
\end{equation}
Observe that we assumed $f(\cdot)$ to be non-decreasing. This combined with \eqref{eq:lamf} and the Bellman equations \eqref{eq:bell1} and \eqref{eq:below_th} ensures that $S(\cdot)$ is also non-decreasing. Thus, if there exists a state $H$ that satisfies \eqref{eq:wc}, then the threshold policy with threshold $H$ satisfies the Bellman equations and is hence optimal. 

The one thing that remains to be shown is the case in which we cannot find some $H$ that satisfies \eqref{eq:wc}. Consider the function $W:\mathbb{Z}^{+}\rightarrow \mathbb{R}$ given by
\begin{equation}
    W(h) = h f(h) - \sum_{j=1}^{h}f(j).
\end{equation}
Observe that $W(h+1) - W(h) = h (f(h+1) - f(h)) \geq 0$ since $f(\cdot)$ is non-decreasing. Thus, $W(\cdot)$ is also non-decreasing. Also, by definition, $W(1)=0$, while we had assumed that $C > 0$. Thus, $W(1)<C$. Now, if there exists some $h>1$ such that $W(h)\geq C$, then we know that there also exists some $H$ such that $W(H) \leq C \leq W(H+1)$ using monotonicity of $W(\cdot)$. Observe that this implies that there exists some $H$ satisfying
\begin{equation*}
    H f\big(H\big) - \sum_{j=1}^{H}f(j) \leq C \leq (H+1)f\big(H+1\big) - \sum_{j=1}^{H+1}f(j).
\end{equation*}
Rearranging and dividing by $H$, we get back \eqref{eq:wc}. Thus, if there exists no $H$ satisfying \eqref{eq:wc}, then $W(h) < C, \forall h$. 

Since $W(\cdot)$ is a bounded monotone sequence, it converges to a finite value. It is easy to see that this implies that $f(\cdot)$ is also bounded and hence converges. We set $\lambda = \lim_{h\rightarrow\infty} f(h)$ and the cost-to-go function $S(h)$ to be
\begin{equation}
\label{eq:sh_inf}
    S(h) = \sum_{j=h}^{\infty} \big(f(j) - \lambda\big) + C.
\end{equation}
Clearly, $S(h)$ satisfies the recurrence relation 
\begin{equation}
    S(h) = f(h) - \lambda + S(h+1), \forall h.
\end{equation}
By the monotonicity of $f(\cdot)$, we know that $f(h)\leq \lambda, \forall h$. Thus, using \eqref{eq:sh_inf} we conclude that $S(h)\leq C, \forall h$. This implies that $S(\cdot)$ satisfies the Bellman equations, with the optimal policy being to never activate the arm. This completes our proof.


\subsection{Proof of Theorem~\ref{thm:rc_rel}}
\label{pf:thm_ind_1}
For $C = 0$, it is obvious that the optimal policy is to always activate the arm since there is no charge for activating it and the cost function is monotone and positive. For larger values of $C$, consider the function $W:\mathbb{Z}^{+}\rightarrow \mathbb{R}^{+}$ given by
\begin{equation}
    W(h) = h f(h) - \sum_{j=1}^{h}f(j).
\end{equation}
Observe that since $f(\cdot)$ is non-decreasing, $W(\cdot)$ is also non-decreasing. This is because $W(h+1) - W(h) = h \big(f(h+1) - f(h)\big) \geq 0, \forall h$ since $f(\cdot)$ is non-decreasing.  Also, by definition, $W(1)=0$, while we had assumed that $C > 0$. Thus, $W(1)<C$. Now, if there exists some $h>1$ such that $W(h)\geq C$, then we know that there also exists some $H$ such that $W(H) \leq C \leq W(H+1)$ using monotonicity of $W(\cdot)$. Observe that this implies that there exists some $H$ satisfying
\begin{equation}
    H f\big(H\big) - \sum_{j=1}^{H}f(j) \leq C \leq (H+1)f\big(H+1\big) - \sum_{j=1}^{H+1}f(j).
\end{equation}
Rearranging and dividing by $H$, we get back \eqref{eq:wc}.

Using this, we can relate the optimal threshold values to values of activation charge. Let $C$ be such that it lies in the interval $\big[W(h),W(h+1)\big)$, then the optimal policy is of threshold type with the threshold at $h$. Observe that if $W$ is strictly increasing then there can only be one such interval in which $C$ can lie. If $W(\cdot)$ is non-decreasing, then there could be multiple such intervals in which $C$ could lie. In this case, we choose the smallest $h$ such that the condition holds. 

The monotonicity of $W(\cdot)$ ensures that the the threshold value is also monotone non-decreasing with increasing values of $C$. When $W(h) < C, \forall h$, we choose $h$ to be $\infty$, as done in Appendix~\ref{pf:thm_wrc}. This completes the proof of \textit{indexability} for the decoupled problem. Observe that $C = W(h+1)$ is the minimum value of the activation charge that makes both actions equally desirable in state $h$. This gives us the expression for the Whittle index.

\subsection{Proof of Theorem~\ref{thm:sw_cyc}}
\label{pf:sw_cyc}

We look at the optimal cyclical policy and analyze its properties. If there are multiple such cycles, we consider the cycle with the shortest length. We denote the length of the cycle by $T$, points on the cycle to be $\bm{x_1},\dots,\bm{x_T}$, and the average cost of this cycle to be $C^{*}$. The point $\bm{x_i}$ is an age vector in $\mathbb{Z}^{+^{N}}$, where $x_{i}^{(j)}$ represents the age of the $j\supth$ source. Let the corresponding scheduling decisions be $d_1,\dots,d_T$. This implies that for age vector $\bm{x_k}$, taking action $d_k$ leads to the age vector $\bm{x_{k+1}}$, where the subscripts cycle back to $1,2,\dots$ after $T$. Assume that there exists some pair of states in this cycle that violate the strong-switch-type property. If not, then our claim that the cycle satisfies the strong-switch-type property is true.

Without loss of generality, we assume that the pair of states that violates strong-switch is given by $\bm{x_1}$ and $\bm{x_k}$ for some $k \in \{ 2,\dots,T \}$. This is because the cyclical policy is same up to cyclical permutations, so we can always ensure that one member of the violating pairs is at the front of the cycle. Also without loss of generality, we assume that $d_1 = 1$ and $d_k = 2$, since we can always relabel the sources. Observe that $d_1$ and $d_k$ cannot be the same since they violate the strong-switch property. In fact, we know that $x_k^{(j)} \leq x_1^{(j)}, \forall j \neq 1$ and $x_k^{(1)} \geq x_1^{(1)}$. If the strong-switch property was satisfied, $d_k$ must have been $1$ since $d_1$ is $1$.

We now construct two new cyclical policies out of which at least one has a better cost or the same cost but a smaller length compared to the original optimal policy. This contradicts our original assumption that the cycle we had started with was the shortest policy with the lowest average cost. Starting with the state $\bm{x_1}$, we take the action $d_1$, following the original cycle up to $\bm{x_k}$. At $\bm{x_k}$, instead of taking action $d_k$, we take the action $d_1$ leading to the state $\bm{y_{k+1}}$. Observe that $\bm{y_{k+1}} \leq \bm{x_{2}}$, where the inequality is element-wise. Since $d_1 = 1$, we schedule source 1 at both $\bm{x_1}$ and $\bm{x_k}$, which guarantees that its age goes to 1. Thus, $x_2^{(1)} = y_{k+1}^{{(1)}} = 1$. Also, since $x_k^{(j)} + 1 \leq x_1^{(j)} + 1, \forall j \neq 1$ and none of the other sources are scheduled, so $y_{k+1}^{(j)}  \leq x_2^{(j)} , \forall j \neq 1.$ Together, this implies $\bm{y_{k+1}} \leq \bm{x_{2}}$. 

Now, we follow the original cycle starting from $d_2, \dots, d_T$. Action $d_2$ at state $\bm{y_{k+1}}$ leads to state $\bm{y_{k+2}}$ and so on, up to action $d_T$ at state $\bm{y_{k+T-1}}$. Since the channels are reliable and $\bm{y_{k+1}} \leq \bm{x_{2}}$, it is easy to see that $\bm{y_{k+i}} \leq \bm{x_{i+1}}, \forall i \in \{1,\dots,T-1\}$. 

Also, observe that starting at $\bm{x_k}$, we have repeated an entire period of the original cycle, i.e. $d_1, \dots,d_T$. Every source gets activated at least once during the original cycle, otherwise, its age goes to infinity, and we might as well remove it from the system. Starting at any age vector and following the actions $d_1, \dots,d_T$ in sequence ensures that the state reached after these $T$ steps equals $\bm{x_1}$. Thus, the actions $\{d_1,\dots,d_{k-1},d_1,\dots,d_T\}$ and the age vectors $\{\bm{y_1},\dots,\bm{y_k},\bm{y_{k+1}},\dots,\bm{y_{k+T-1}}\}$ form a cycle of length $k+T-1$. Here $\bm{y_i} = \bm{x_i}, \forall i \in 1,\dots,k$ and $\bm{y_{k+i}} \leq \bm{x_{i+1}}, \forall i \geq 1$. We denote the average cost of this cycle by $C_1$.

Now, we perform a cyclic permutation of the original optimal policy to get a new optimal policy with the actions $\{d_k, \dots,d_T,d_1,\dots,d_{k-1}\}$ and the corresponding states $\{ \bm{x_{k}},\dots,\bm{x_T},\bm{x_1},\dots,\bm{x_{k-1}} \}$. We repeat the process of constructing a new cyclical policy of length $2T-k+1$ as done above, but using the new cyclic permutation of the optimal policy. That is, instead of choosing action $d_1$ at $\bm{x_k}$, we choose action $d_k$ at $\bm{x_1}$.

This new cyclical policy consists of actions $\{d_k,\dots,d_{T}, d_k,\dots,d_T,d_1,\dots,d_{k-1}\}$ and the corresponding age vectors $\{\bm{z_1},\dots,\bm{z_{2T-k+1}}\}$, forming a cycle of length $2T-k+1$. Using exactly the same argument as earlier, it is easy to see that $\bm{z_{j}} = \bm{x_{j+k-1}}, \forall j \in \{1,\dots,T-k+1\}$ and $\bm{z_{j}} \leq \bm{x_{j}}, \forall j \in \{T-k+2,\dots,2T-k+1\}$. We denote the average cost of this cycle by $C_2$.

We know that the cost of the optimal policy $C^{*}$ is minimum over the space of all policies, and hence less than or equal to cost of the first cyclical policy that we created $C_1$. Thus,
\begin{equation*}
\begin{split}
    C^{*} &\leq C_1\\
    \implies \frac{1}{T}\sum_{t=1}^{T} &\sum_{j=1}^{N}f_j(x_t^{(j)}) \leq  \frac{1}{k+T-1}\sum_{t=1}^{k+T-1} \sum_{j=1}^{N}f_j(y_t^{(j)})\\
     \leq \frac{1}{k+T-1} &\bigg(\sum_{t=1}^{k} \sum_{j=1}^{N}f_j(x_t^{(j)}) + \sum_{t=k+1}^{k+T-1} \sum_{j=1}^{N}f_j(y_t^{(j)})\bigg)\\
     \leq \frac{1}{k+T-1} &\bigg(\sum_{t=1}^{k} \sum_{j=1}^{N}f_j(x_t^{(j)}) + \sum_{t=2}^{T} \sum_{j=1}^{N}f_j(x_t^{(j)})\bigg)
\end{split}
\end{equation*}
Simplifying this inequality, we get
\begin{equation}
\label{eq:sw_con1}
   \frac{1}{T}\sum_{t=1}^{T} \sum_{j=1}^{N}f_j(x_t^{(j)}) \leq \frac{1}{k-1}\sum_{t=2}^{k} \sum_{j=1}^{N}f_j(x_t^{(j)}). 
\end{equation}
Similarly, the average cost of the optimal cycle $^{*}$ is also less than or equal to the average cost of the second cycle $C_2$. Thus,
\begin{equation*}
\begin{split}
    &C^{*} \leq C_2\\
    \implies &\frac{1}{T}\sum_{t=1}^{T} \sum_{j=1}^{N}f_j(x_t^{(j)}) \leq  \frac{1}{2T-k+1}\sum_{t=1}^{2T-k+1} \sum_{j=1}^{N}f_j(z_t^{(j)})\\
     &\leq \frac{1}{2T-k+1}\bigg(\sum_{t=k}^{T} \sum_{j=1}^{N}f_j(x_t^{(j)}) + \sum_{j=1}^{N}f_1(x_1^{(j)}) + \\
     &\sum_{t=T-k+3}^{2T-k+1} \sum_{j=1}^{N}f_j(z_t^{(j)})\bigg)\\
     &\leq \frac{1}{2T-k+1} \bigg(\sum_{t=k+1}^{T} \sum_{j=1}^{N}f_j(x_t^{(j)})  + \sum_{j=1}^{N}f_1(x_1^{(j)}) + \\
     & \sum_{t=1}^{T} \sum_{j=1}^{N}f_j(x_t^{(j)})\bigg)
\end{split}
\end{equation*}
Simplifying this inequality, we get
\begin{equation*}
\begin{split}
   \frac{1}{T}\sum_{t=1}^{T} \sum_{j=1}^{N}f_j(x_t^{(j)}) \leq \frac{1}{T-k+1}\bigg(\sum_{t=k+1}^{T} \sum_{j=1}^{N}f_j(x_t^{(j)}) \\+ \sum_{j=1}^{N}f_1(x_1^{(j)})\bigg).
\end{split}
\end{equation*}
Rearranging and simplifying again, we get
\begin{equation}
\label{eq:sw_con2}
   \frac{1}{T}\sum_{t=1}^{T} \sum_{j=1}^{N}f_j(x_t^{(j)}) \geq \frac{1}{k-1}\sum_{t=2}^{k} \sum_{j=1}^{N}f_j(x_t^{(j)}). 
\end{equation}

From the analysis above, we observe that \eqref{eq:sw_con1} and \eqref{eq:sw_con2} must hold simultaneously. However, if that's the case then the inequalities cannot be strict. Also, observe that the cyclical policy given by actions $\{d_2,...,d_{k}\}$ has average cost $C_3$ that satisfies
\begin{equation*}
    C_3 \leq \frac{1}{k-1}\sum_{t=2}^{k} \sum_{j=1}^{N}f_j(x_t^{(j)}).
\end{equation*}
This is because starting at state $\bm{x_2}$ and following the policy we end up at state $\bm{x_k}$, where using the exact same argument as earlier, taking action $d_k$ leads us to a state $\bm{y_{k+1}}$ such that $\bm{y_{k+1}}\leq x_2$. The upper bound follows directly. Also, since \eqref{eq:sw_con1} is tight, we get that 
\begin{equation*}
    C_3 \leq \frac{1}{T}\sum_{t=1}^{T} \sum_{j=1}^{N}f_j(x_t^{(j)}).
\end{equation*}
This is a contradiction, since if the above inequality is strict, our original policy is no longer optimal and if the inequality is tight, we have a smaller length cycle with the same cost, which still contradicts our original assumption that we started with an optimal cost cycle with minimum length.

\subsection{Proof of Theorem~\ref{thm:sw_opt}}
\label{pf:sw_opt}

We have shown that the points on the optimal cycle satisfy the strong-switch-type property. We need to show that we can assign actions to states that are not on the optimal cycle while maintaining the strong-switch property, for $N \leq 3$.
\color{black}

This can be done in an iterative manner. Consider the set of points in the state-space that have been assigned an action, and which satisfy the strong-switch property to be $D$. Let $\bm{x} \notin D$, be a new point that we want to assign an action to. There are three possible scenarios - 1)  there exists $\bm{y} \in D$ such that the strong-switch-type property implies a unique action to be taken at $\bm{x}$, 2) there exists no such $\bm{y} \in D$ and so an arbitrary action can be chosen at $\bm{x}$, and 3) there exist multiple such points in $D$, which suggest different actions to be taken at $\bm{x}$. 

Clearly, for scenarios 1 and 2 above, we can assign an action to the point $\bm{x}$, increase our set to $D \cup \{\bm{x}\}$ and repeat the procedure for a new point. We claim that if $N \leq 3$ then scenario 3 never occurs. This is sufficient to prove that we can extend the strong-switch-type property over the entire state-space. 

To prove that scenario 3 doesn't happen, we start by assuming the contrary. Let $\bm{y_1} \text{ and } \bm{y_2} \in D$ and without loss of generality, assume that the action taken at $\bm{y_1}$ is 1 and the action taken at $\bm{y_2}$ is 2. Also, to satisfy our assumption of scenario 3 for $\bm{x}$, we require that $x^{(1)} \geq y^{(1)}_1$, $x^{(2)} \geq y^{(2)}_2$, $x^{(j)} \leq y^{(j)}_1, \forall j \neq 1$ and $x^{(k)} \leq y^{(k)}_1, \forall k \neq 2$. For these inequalities to be feasible simultaneously, we need $y^{(1)}_2 \geq y^{(1)}_1$ and $y^{(2)}_1 \geq y^{(2)}_2$.

Now, if there are only two sources, i.e. $N=2$, then the fact that $y^{(1)}_2 \geq y^{(1)}_1$ and $y^{(2)}_1 \geq y^{(2)}_2$ together with the assumption that the action taken at $\bm{y_1}$ is 1 and the action taken at $\bm{y_2}$ is 2, we get that $\bm{y_1}$ and $\bm{y_2}$ violate the strong-switch property, despite being in the set $D$. This is a contradiction and completes our proof.

Similarly, consider the setting with three sources (N=3). Now, there are two possibilities - either $y_1 ^{(3)} \leq y_2 ^{(3)}$ or $y_1 ^{(3)} > y_2 ^{(3)}$. If $y_1 ^{(3)} \leq y_2 ^{(3)}$, then using the fact that $y^{(1)}_1 \leq y^{(1)}_2$ and $y^{(2)}_1 \geq y^{(2)}_2$, the strong-switch-type property implies that the action taken at $\bm{y_1}$ must be $2$. However, we assumed that the action taken at state $\bm{y_1}$ is 1. Thus, this violates the strong-switch-type property. Similarly, if $y_1 ^{(3)} > y_2 ^{(3)}$, then using the fact that $y^{(1)}_1 \leq y^{(1)}_2$ and $y^{(2)}_1 \geq y^{(2)}_2$, the strong-switch-type property implies that the action taken at $\bm{y_2}$ must be $1$. This again violates our assumption that $\bm{y_1}$ and $\bm{y_2}$ satisfy the strong-switch-type property. 

Thus, we have proved that for $N \leq 3$, if the optimal cycle is strong-switch-type then we can find a stationary optimal policy that is strong-switch-type over the entire state-space.



\subsection{Proof of Theorem \ref{thm:eq_pol}}
\label{pf:thm_eq_pol}
We use an inductive argument to prove this result. Assume that for a scheduling setup with $N-1$ sources, every strong-switch type policy can also be written as an index policy. Now, consider a functions of age setup with $N$ sources and reliable channels. Using Theorem \ref{thm:sw_opt} we know that there exists an optimal policy that is strong-switch-type. Let this policy be $\pi:\mathbb{Z}^{+^{N}}\rightarrow\{1,\dots,N\}$. 

Let $x_i$ denote the age of the $i\supth$ source when the current state is $\bm{x}$ and let $\bm{x_{-i}}$ denote the vector comprising of ages of all sources except $i$. Consider the minimum age $x_1$ at source 1 such that $\pi(x_1,\bm{x_{-1}}) = 1$, for any fixed $\bm{x_{-1}}$. That is, for any fixed value of ages for all other sources, consider the age at source $1$ for which the optimal policy schedules the first source. This value of $x_1$ may depend on $\bm{x_{-1}}$, so we denote it by $x_{\text{th}}(\bm{x_{-1}})$. Observe that for all values of $x_1$ such that $x_1 \geq x_{\text{th}}(\bm{x_{-1}})$, the strong-switch-type property implies that $\pi(x_1,\bm{x_{-1}}) =1$. In other words, $x_{\text{th}}(\bm{x_{-1}})$ acts like a threshold value such that for all values of age at source $1$ above it, the optimal policy schedules the first source. If no such threshold exists, we let $x_{\text{th}}(\bm{x_{-1}}) \rightarrow \infty$. 

We append the state space of the first arm by zero, i.e. let the state space of source $1$ be $\mathbb{Z}_0 = \mathbb{Z}^+ \cup \{0\}$. Zero is the minimum age that this source can have and without loss of generality, we can set $f_1(0) = 0$. If at any time-slot the age of this source is zero, we let it increases to one in the next time-slot. Scheduling this source when its age is zero gives us no benefit, as the age increases by one no matter what our scheduling decision is. We extend the policy $\pi$ over this new state space as follows. Let $\pi':\mathbb{Z}_0\times\mathbb{Z}^{+^{N-1}}\rightarrow\{1,\dots,N\}$ be a mapping that satisfies
\begin{itemize}
    \item $\pi'(\bm{x}) = \pi(\bm{x}), \forall \bm{x} \text{ s.t. }x_1 \neq 0$.
    \item if $x_{\text{th}}(\bm{x_{-1}}) > 1$, $\pi'(0,\bm{x_{-1}}) = \pi(x_{\text{th}}(\bm{x_{-1}})-1,\bm{x_{-1}})$
    \item if $x_{\text{th}}(\bm{x_{-1}}) = 1$, $\pi'(0,\bm{x_{-1}}) = \pi_{N-1}(\bm{x_{-1}})$,
\end{itemize}
where $\pi_{N-1}(\bm{x_{-1}})$ is an optimal strong-switch-type policy for the functions of age problem with just the sources $2,\dots,N$. It is easy to see that this new extended policy still satisfies the strong-switch-type property and is still optimal for the original problem with all $N$ sources over the extended state space.

Now, we project this new optimal policy $\pi'$ on to $\mathbb{Z}^{+^{N-1}}$ to get a new policy $\pi'':\mathbb{Z}^{+^{N-1}}\rightarrow\{2,\dots,N\}$ such that $\pi''(\bm{x_{-1}}) = \pi'(x_{\text{th}}(\bm{x_{-1}})-1,\bm{x_{-1}})$. This is well defined since $x_{\text{th}}(\bm{x_{-1}}) \geq 1$ and $\pi'(x_{\text{th}}(\bm{x_{-1}})-1,\bm{x_{-1}}) \in \{2,\dots,N\}$, by construction. Also, $\pi''$ is strong-switch-type by construction, since it is a projection of a strong-switch-type policy onto a lower dimensional space. If not, then $\pi'$ would also violate the strong-switch-type property.

Now, using our induction assumption, we can find index functions such that 
\begin{equation*}
    \pi''(x_2,\dots,x_N) = \text{arg}\underset{2 \leq i \leq N}{\text{max}} \bigg\{ F_i(x_i) \bigg\}
\end{equation*}
for all $\bm{x}$, where $F_i:\mathbb{Z}^{+}\rightarrow \mathbb{R}$ are monotonically non-decreasing functions for all $i$.

We partition the $N-1$ dimensional state space of policy $\pi''$ into a countable number of sets. Let
\begin{equation*}
    S_k \triangleq \{\bm{x}:\bm{x} \in \mathbb{Z}^{+^{N-1}}, x_{\text{th}}(\bm{x}) = k \}, \forall k \in \mathbb{Z}^{+}.
\end{equation*}
Then, $\mathbb{Z}^{+^{N-1}} = \cup_{k=1}^{\infty} S_k$ and $S_k \cap S_j = \{\phi\}, \forall k,j$. Consider $\bm{x} \in S_j$ and $\bm{y} \in S_k$ such that $k > j$. Then,
\begin{equation}
\label{eq:ind_con1}
   \underset{2 \leq i \leq N}{\text{max}} \bigg\{ F_i(x_i) \bigg\} \leq \underset{2 \leq i \leq N}{\text{max}} \bigg\{ F_i(y_i) \bigg\}.
\end{equation}
Suppose the opposite is true, i.e. $\underset{2 \leq i \leq N}{\text{max}} \bigg\{ F_i(x_i) \bigg\} > \underset{2 \leq i \leq N}{\text{max}} \bigg\{ F_i(y_i) \bigg\}$. Let $m = \text{arg}\underset{2 \leq i \leq N}{\text{max}} \bigg\{ F_i(x_i) \bigg\}$. Clearly, for the opposite of \eqref{eq:ind_con1} to hold we need $x_m > y_m$. If we define $\bm{z}$ such that $z_i \triangleq \max\{ x_i,y_i\}, \forall i  \in 2,\dots,N$, then using the index property of $\pi''(\cdot)$ we get
\begin{equation*}
    \pi''(\bm{z}) = m.
\end{equation*}
Also, $x_{\text{th}}(\bm{z}) \geq k$. If not, then since $\bm{z} \geq \bm{y}$, the strong-switch property implies $\pi'(x_{\text{th}}(\bm{z}),\bm{y})=1$, where $x_{\text{th}}(\bm{z}) < k$. This violates our assumption that $\bm{y} \in S_k$.

Now, observe that $\pi'(x_{\text{th}}(\bm{z})-1,\bm{z}) = m$, since $\pi''(\bm{z}) = m$. Also $(j, x_2,\dots,x_m,\dots,x_{N})\leq (x_{\text{th}}(\bm{z})-1,z_2,\dots,x_m,\dots,z_{N})$ where the inequality holds element-wise. This is because $z_i = \max\{ x_i,y_i\}$, and $j \leq k-1 \leq x_{\text{th}}(\bm{z})-1$, and $x_m > y_m$. Thus, using the strong-switch-type property of $\pi'(\cdot)$, we get
\begin{equation*}
    \pi'(j,\bm{x}) = m.
\end{equation*}
This contradicts our initial assumption that $x \in S_j$, since that would imply $\pi'(j,\bm{x}) = 1.$ Thus, we conclude that \eqref{eq:ind_con1} must be satisfied.

We now construct a monotone function based on the above discussion. Let
\begin{equation}
    F_1(j) \triangleq \sup_{\bm{z} \in S_j} \underset{2 \leq i \leq N}{\text{max}} \bigg\{ F_i(z_i) \bigg\}, \forall j \in \mathbb{Z}^{+}.
\end{equation}
Clearly, since the condition \eqref{eq:ind_con1} is satisfied, the function $F_1(\cdot)$ is monotone. Also, let 
\begin{equation*}
    \pi'''(\bm{x}) = \text{arg }\underset{1 \leq i \leq N}{\text{max}} \bigg\{ F_i(x_i) \bigg\}, \forall \bm{x} \in \mathbb{Z}^{+^{N}},
\end{equation*}
where we break ties in lexicographic order. Then $\pi'''$ is the same our original policy $\pi$. This is because for every state $\bm{x} \in S_j$, the construction of $F_1$ forces us to schedule source $1$ for values of $x_1 \geq j$ and not schedule source $1$ for values below $j$. This holds for all values of $j$, which means we replicate the original scheduling policy $\pi(\cdot)$. Thus, if we assume strong-switch-type policies can be written as index policies for a problem with $N-1$ sources, we can also prove the same fact for $N$ sources.

It is trivial to see that strong-switch-type policies and index policies are equivalent for the single source decoupled problem. This is because strong-switch-type policies and index policies both correspond to monotone threshold policies for the decoupled problem. Hence, using the principle of induction, we have the required result.

\subsection{Proof of Theorem \ref{thm:2_opt}}
\label{pf:thm_2_opt}

Using Corollary \ref{corr_opt}, we know that there exists some index policy which is optimal.We observe that for $N=2$ index policies have a specific structure. 

Let $F_1(\cdot)$ and $F_2(\cdot)$ represent the index functions for the optimal index policy.  We set the ages of the two sources to $(1,1)$ at time $t=1$ and assume that the optimal index functions are such that $F_1(1) \geq F_2(1)$. Then, the policy schedules source $1$ at time $t=1$. The new state at time $t=2$ is given by $(1,2)$. Again, assume that $F_1(1) \geq F_2(2)$. Then, the policy schedules source $1$ at time $t=2$. The new state at time $t=3$ is given by $(1,3)$. We keep repeating this process until we reach state $(1,k)$ at time $t=k$ for which $F_1(1) < F_2(k)$. The policy then schedules source $2$ and reaches state $(2,1)$ at time $k+1$. Now, since we assumed that $F_1(1)\geq F_2(1)$, then using monotonicity we get $F_1(2)\geq F_2(1)$. Thus, the policy schedules source $1$ again and we reach state $(1,2)$ at time $t=k+2$.

From the above discussion, we see that any index policy for $N=2$ has a cyclic form and the cycle consists one of the sources being scheduled repeatedly followed by the second source once. To find the best index policy, which is also the best policy overall, we just need to find the best policy with this specific structure.

Without loss of generality, assume that an optimal cyclical policy is given by scheduling source 1 $k$ times followed by source $2$ once, and repeating this sequence of actions. Now, consider two cases.
\subsubsection{Case 1} $(k>1)$ We compare the cost of the optimal cycle with a cycle that schedules source 1 $k-1$ times followed by source 2 once.  
\begin{equation*}
\begin{split}
    \frac{\sum_{j=1}^{k} f_2(j) + (k-1)f_1(1) + f(2)}{k} \geq\\ \frac{\sum_{j=1}^{k+1} f_2(j) + kf_1(1) + f(2)}{k+1}
\end{split}
\end{equation*}
Simplifying, we get
\begin{equation*}
  f_1(2) - f_1(1) \geq k f_2(k+1) - \sum_{j=1}^{k} f_2(j),
\end{equation*}
i.e. $W_1(1) \geq W_2(k)$. The Whittle index policy follows the optimal policy till the state $(1,k)$.

We then compare the cost of the optimal cycle with a cycle that schedules source 1 $k+1$ times followed by source 2 once.  
\begin{equation*}
\begin{split}
    \frac{\sum_{j=1}^{k+2} f_2(j) + (k+1)f_1(1) + f(2)}{k+2} >\\ \frac{\sum_{j=1}^{k+1} f_2(j) + kf_1(1) + f(2)}{k+1}
\end{split}
\end{equation*}
Simplifying, we get
\begin{equation*}
  (k+1)f_2(k+2) - \sum_{j=1}^{k+1} f_2(j) > f_1(2) - f_1(1),
\end{equation*}
i.e. $W_2(k+1) > W_1(1)$.

Together, this implies that the Whittle Index policy must also schedule source 1 $k$ times followed by source 2 once, and repeat this sequence of actions. Hence, the Whittle index policy is optimal.

\subsubsection{Case 2} $(k = 1)$ We compare the optimal policy with a cycle that schedules source $1$ twice and source 2 once. Then, we get
\begin{equation*}
\begin{split}
    \frac{2f_1(1)+f_1(2) + f_2(1)+f_2(2)+f_2(3)}{3} > \\ \frac{f_1(1) + f_2(1) + f_1(2) + f_2(2)}{2}
\end{split}
\end{equation*}
Simplifying, we get 
\begin{equation*}
    2f_2(3) - f_2(1) - f_2(2) > f_1(2) - f_1(1),
\end{equation*}
i.e. $W_2(2) > W_1(1)$. Using a symmetrical argument, it is easy to see that $W_1(2) > W_2(1)$. Thus, the Whittle Index policy also schedules each source exactly once, and repeats this sequence of actions.

Combining the two cases, we conclude that for $N=2$ and reliable channels, the Whittle Index policy is exactly optimal.

\subsection{Proof of Theorem \ref{thm:wurc}}
\label{pf:thm_wurc}

Let $S:\mathbb{Z}^{+}\rightarrow\mathbb{R}$ denote the differential cost-to-go function for this problem, let $u:\mathbb{Z}^{+}\rightarrow \{1,0 \}$ be the stationary optimal policy and let $\lambda$ denote the optimal cost. Then, the Bellman equations are given by
 \begin{equation}
 \begin{split}
     S(h) = f(h) + \underset{u(h) \in \{1,0\}}{\text{min}} \{C + (1-p)S(h+1),S(h+1)\} \\- \lambda, \forall h \in \mathbb{Z}^{+}.
 \end{split}
 \end{equation}
Without loss of generality we set $S(1) = 0$. Assume that the optimal policy has a threshold structure, i.e. there exists $H$ such that it is optimal to pull the arm $(u(h)=1)$ for all states $h \geq H$ and let it rest otherwise $(u(h)=0)$. If this the case, then the Bellman equations for values above the threshold $H$ reduce to
\begin{equation}
\label{eq:bell1_urc}
    \begin{split}
    S(h) & = f(h) + C + (1-p)S(h+1)- \lambda, \forall h \geq H.
    \end{split}
\end{equation}
Solving this recursion and assuming $\lim_{h \rightarrow \infty} (1-p)^{h}S(h) = 0$, we get
\begin{equation}
\label{bells_urc}
    S(H + j) = \sum_{k = j}^{\infty} f(k+H) (1-p)^{k-j} + \frac{C-\lambda}{p}, \forall j \geq 0.
\end{equation}
Since $f(\cdot)$ is non-decreasing, it is easy to see that $S(h)$ is also non-decreasing for all values of $h$ above the threshold $H$, using \eqref{bells_urc}. We will use this fact later. Now, observe that 
\begin{equation}
\begin{split}
\lim_{h \rightarrow \infty} (1-p)^{h}S(h) &= \lim_{h \rightarrow \infty} \sum_{j=h}^{\infty} f_i(j) (1-p_i)^{j} \\ &+ \lim_{h \rightarrow \infty}  \frac{C - \lambda}{h}(1-p_i)^{h}, \forall h \geq H.
\end{split}
\end{equation}
By the bounded cost assumption, the first term is the limit of the partial sums of a convergent series, thus it goes to zero. The second term also goes to zero since $p<1$ and $\lambda$ is finite, again using the bounded cost assumption. This confirms that our assumption $\lim_{h \rightarrow \infty} (1-p)^{h}S(h) = 0$ was indeed correct. 

For $h = H-1$, the Bellman equation is given by
\begin{equation}
\begin{split}
S(H&-1) = f\big(H-1\big) - \lambda + S(H)\\
      & = f\big(H-1\big) - \lambda + \sum_{k = 0}^{\infty} f(k+H) (1-p)^{k} + \frac{C-\lambda}{p}.
\end{split}
\end{equation}
Repeating this $k$ times, we get
\begin{equation}
\label{eq:bel_th_urc}
\begin{split}
    S(H-k) = \sum_{j=H-k}^{H-1} (f(j) - \lambda) + \sum_{j = 0}^{\infty} f(j+H) (1-p)^{j} \\+ \frac{C-\lambda}{p}, \forall k \in \{1,\dots,H-1\}. 
\end{split}
\end{equation}
Now, putting $k = H-1$ in the above equation and using the fact that $S(1) = 0$, we get 
\begin{equation}
\label{eq:lam2_urc}
    \lambda = \frac{p\big(\sum_{j=1}^{H}f(j) + \sum_{k=1}^{\infty} f(k+H)(1-p)^{k}\big)  + C }{1 + p(H-1)}.
\end{equation}
If we further assume that the threshold value $H$ satisfies the condition \eqref{eq:urc_con} given in Theorem~\ref{thm:wurc}, then we get that
\begin{equation}
\begin{split}
   p^2 (H-1) &\bigg(\sum_{k=H}^{\infty} f(k)(1-p)^{k-H}\bigg) - p \bigg(\sum_{j=1}^{H-1}f(j)\bigg) \\ \leq &~C\\ \leq p^2 H &\bigg(\sum_{k=H+1}^{\infty} f(k)(1-p)^{k-H-1}\bigg) - p \bigg(\sum_{j=1}^{H}f(j)\bigg).
\end{split}
\end{equation}
Rearranging terms, dividing by $1+p(H-1)$ and using the expression for $\lambda$ from \eqref{eq:lam2_urc}, we get
\begin{equation}
\label{eq:lamf_urc}
\begin{split}
   p\bigg(\sum_{k=0}^{\infty} f(H+k)(1-p)^{k}\bigg)  \leq \lambda \\ \leq  p\bigg(\sum_{k=1}^{\infty} f(H+k)(1-p)^{k-1}\bigg).
\end{split}
\end{equation}
Simplifying the inequalities in \eqref{eq:lamf_urc}, the expression for $\lambda$ from \eqref{eq:lam2_urc} and the Bellman solutions \eqref{bells_urc}, we get 
\begin{equation}
\label{eq:dp_urc}
    S(H) \leq \frac{C}{p} \leq S(H+1).
\end{equation}
Using \eqref{eq:bel_th_urc}, we note that for $h < H$, $S(h) - S(h-1) = \lambda - f(h-1)$. Also, using the monotonicity of $f(\cdot)$ and \eqref{eq:lamf_urc}, we get
\begin{equation}
\begin{split}
   \lambda &\geq  p \bigg(\sum_{k=0}^{\infty} f(H+k)(1-p)^{k}\bigg)  \\
   &\geq  p \bigg(\sum_{k=0}^{\infty} f(H)(1-p)^{k}\bigg) \\
   &\geq f(H)
   \\& \geq f(h), \forall h < H.
\end{split}
\end{equation}

Thus, $S(h) - S(h-1) \geq 0, \forall h$ since we already established monotonocity for $h\geq H$. Since $S(\cdot)$ is non-decreasing, \eqref{eq:dp_urc} implies that 
\begin{equation}
    \begin{split}
        S(h) &\leq C + (1-p)S(h), \forall h \leq H, \text{ and} \\
        S(h) &\geq C + (1-p)S(h), \forall h > H.
    \end{split}
\end{equation}
Thus, if we find an $H$ that satisfies \eqref{eq:urc_con}, the threshold policy using $H$ as a threshold satisfies the Bellman equations and is optimal. 

The one thing that remains to be shown is the case in which we cannot find some $H$ that satisfies \eqref{eq:wc}. As done earlier, we define a function $W:\mathbb{Z}^{+}\rightarrow \mathbb{R}$ given by
\begin{equation}
    W(h) = p^2 (h-1) \bigg(\sum_{k=h}^{\infty} f(k)(1-p)^{k-h}\bigg) - p \bigg(\sum_{j=1}^{h-1}f(j)\bigg).
\end{equation}
Observe that 
\begin{equation}
\begin{split}
    W&(h+1) - W(h) =\\
    &p^2 h \bigg[ \sum_{k=0}^{\infty} \big(f(h+1+k) - f(h+k)\big)(1-p)^{k} \bigg] \\
    &+ p^2 \bigg[ \sum_{k=0}^{\infty}f(h+k)(1-p)^{k} \bigg] - pf(h)\\
    &\geq 0, \forall h
\end{split}
\end{equation}
since $f(\cdot)$ is non-decreasing. Thus, $W(\cdot)$ is also non-decreasing. Also, putting $h=1$ in the definition of $W(h)$ we get $W(1)=0$, while we had assumed that $C > 0$. Thus, $W(1)<C$. Now, if there exists some $h>1$ such that $W(h)\geq C$, then we know that there also exists some $H$ such that $W(H) \leq C \leq W(H+1)$ using monotonicity of $W(\cdot)$. Observe that this implies that there exists some $H$ satisfying \eqref{eq:urc_con} and hence the threshold policy is optimal. If there exists no $H$ satisfying \eqref{eq:urc_con}, then $W(h) < C, \forall h$. 

Since $W(\cdot)$ is a bounded monotone sequence, it converges to a finite value. It is easy to see that this implies that $f(\cdot)$ is also bounded and hence converges. We set $\lambda = \lim_{h\rightarrow\infty} f(h)$ and the cost-to-go function $S(h)$ to be
\begin{equation}
\label{eq:sh_inf2}
    S(h) = \sum_{j=h}^{\infty} \big(f(j) - \lambda\big) + C.
\end{equation}
Clearly, $S(h)$ satisfies the recurrence relation 
\begin{equation}
    S(h) = f(h) - \lambda + S(h+1), \forall h.
\end{equation}
By the monotonicity of $f(\cdot)$, we know that $f(h)\leq \lambda, \forall h$. Thus, using \eqref{eq:sh_inf2} we conclude that $S(h)\leq C, \forall h$. This implies that $S(\cdot)$ satisfies the Bellman equations, with the optimal policy being to never activate the arm. This completes our proof.

\subsection{Proof of Theorem \ref{thm:urc_rel}}
\label{pf:thm_urc_rel}
This proof is very similar to the indexability proof for the reliable channels case. For $C = 0$, it is obvious that the optimal policy is to always activate the arm since there is no charge for activating it and the cost function is monotone and positive. For larger values of $C$, consider the function $W:\mathbb{Z}^{+}\rightarrow \mathbb{R}^{+}$ given by
\begin{equation*}
    W(h) = p^2 (h-1) \bigg(\sum_{k=h}^{\infty} f(k)(1-p)^{k-h}\bigg) - p \bigg(\sum_{j=1}^{h-1}f(j)\bigg).
\end{equation*}
Observe that since $f(\cdot)$ is non-decreasing, $W(\cdot)$ is also non-decreasing, as discussed in Appendix \ref{pf:thm_wurc}.  Also, by definition, $W(1)=0$, while we had assumed that $C > 0$. Thus, $W(1)<C$. Now, if there exists some $h>1$ such that $W(h)\geq C$, then we know that there also exists some $H$ such that $W(H) \leq C \leq W(H+1)$ using monotonicity of $W(\cdot)$. Observe that this implies that there exists some $H$ satisfying \eqref{eq:urc_con}.

Using this, we can relate the optimal threshold values to values of activation charge. Let $C$ be such that it lies in the interval $\big[W(h),W(h+1)\big)$, then the optimal policy is of threshold type with the threshold at $h$. Observe that if $W$ is strictly increasing then there can only be one such interval in which $C$ can lie. If $W(\cdot)$ is non-decreasing, then there could be multiple such intervals in which $C$ could lie. In this case, we choose the smallest $h$ such that the condition holds. 

The monotonicity of $W(\cdot)$ ensures that the the threshold value is also monotone non-decreasing with increasing values of $C$. When $W(h) < C, \forall h$, we choose $h$ to be $\infty$, as done in Appendix~\ref{pf:thm_wurc}. This completes the proof of \textit{indexability} for the decoupled problem. Observe that $C = W(h+1)$ is the minimum value of the activation charge that makes both actions equally desirable in state $h$. This gives us the expression for the Whittle index.

\subsection{Proof of Theorem \ref{thm:lti_fAoI}}
\label{pf:lti_fAoI}
Suppose that the base station knows that the $i$th process was at state $x_i(\tau)=x_0$ at time $\tau$. Further, suppose that it received no additional updates regarding the $i$th process up to time-slot $\tau+\Delta$. Without loss of generality, we can set $\tau=0$, since we can always offset the time-slots by a fixed constant. 

Then, using the state evolution equation \ref{eq:lti_evolution}, we know that
\begin{equation}
	\begin{aligned}
		x_i(1) &= G_i x_0 + w_i(0)\\
		x_i(2) &= G^2_i x_0 + G_i w_i (0) + w_i(1)\\
		...\\
		x_i(\Delta) &= G^{\Delta}_i x_0 + \sum_{k=0}^{\Delta-1} G^{\Delta-k-1}_i w_i(k). 
	\end{aligned}
\end{equation} 

The base station does not have access to the increments $w_i(0),...,w_i(\Delta-1)$. However, it knows that each of them is i.i.d. and $\mathcal{N}(0,\Sigma_i)$. Thus, the maximum likelihood of the state at time $\Delta$ is given by
\begin{equation}
	\hat{x}_i(\Delta) = \mathbb{E}[ x_i(\Delta) | x_i(0) =x_0 ] =  G^{\Delta}_i x_0.
\end{equation}
Using this, we can now compute the difference between the actual state and the estimate at the base station
\begin{equation}
	x_i(\Delta) - \hat{x}_i(\Delta) = \sum_{k=0}^{\Delta-1} G^{\Delta-k-1}_i w_i(k).
\end{equation}
Observe that this is simply a sum of zero-mean independent multi-variate normal random variables. Thus, $x_i(\Delta) - \hat{x}_i(\Delta)$ is also a zero-mean multi-variate normal random variable. 

Recall the following standard properties of multivariate normal random variables. If $X\sim \mathcal{N}(0,\Sigma)$ and $Y=GX$ is some linear transformation of $X$, then $Y\sim\mathcal{N}(0,G\Sigma G^T)$. Further, if $X_1\sim \mathcal{N}(0,\Sigma_1)$ and $X_2\sim \mathcal{N}(0,\Sigma_2)$ are independent, then $Z=X_1 + X_2$ is distributed as $\mathcal{N}(0,\Sigma_1 + \Sigma_2)$. Finally, if $X\sim \mathcal{N}(0,\Sigma)$, then $\mathbb{E}[X^T X ] = Tr(\Sigma)$.

Putting the first two properties together, we observe that
\begin{equation}
	x_i(\Delta) - \hat{x}_i(\Delta) \sim \mathcal{N}\bigg( 0, \sum_{k=0}^{\Delta-1} G^{k}_i \Sigma_i  (G^{k}_i)^T \bigg).
\end{equation}
Using the last property, we get
\begin{equation}
	\begin{aligned}
		e_i(\Delta) &= \mathbb{E}\big[ (x_i(\Delta) - \hat{x}_i(\Delta))^T (x_i(\Delta) - \hat{x}_i(\Delta)) \big]\\
		&=  \mathbb{E}\big[  || x_i(\Delta) - \hat{x}_i(\Delta) ||^2_2 \big]\\
		&= Tr \bigg( \sum_{k=0}^{\Delta-1} G^{k}_i \Sigma_i  (G^{k}_i)^T \bigg)\\
		&= \sum_{k=0}^{\Delta-1} Tr\big((G_i^k)  \Sigma_i  (G_i^k)^T \big)\\
		&= \sum_{k=0}^{\Delta-1} Tr\big((G_i^k)^T (G_i^k) \Sigma_i\big) \triangleq f_i(\Delta).
	\end{aligned}
\end{equation}
The last two equalities follow from the linearity of the trace operator and the fact that $Tr(AB) = Tr(BA)$. This completes the proof of Theorem \ref{thm:lti_fAoI}.

We also want to show that $ f_i(\Delta)$ increases monotonically in $\Delta$. This is straightforward to show, since $G^{k}_i \Sigma_i  (G^{k}_i)^T$ is a covariance matrix for any $k\in \mathbb{Z}^+$. This implies that it must be positive semi-definite, and in turn, must have a non-negative trace. Now, we consider the difference
\begin{equation}
	f_i(\Delta+1) - f_i(\Delta) = Tr\big( (G_i^\Delta) \Sigma_i (G_i^\Delta)^T \big) \geq 0.
\end{equation}
The last inequality follows due to the non-negativity of trace for a positive semi-definite matrix. This shows that $f_i(\Delta+1) \geq f_i(\Delta)$, which allows us to conclude monotonicity of the AoI cost functions.

\subsection{Proof  of Theorem \ref{thm:markov_fAoI}}
\label{pf:markov_fAoI}
As we discussed earlier, the estimate distribution at time $\tau+\Delta$, given the last observation at time $\tau$ is $x_i(\tau)$, is given by $$\hat{x}_i(\tau+\Delta) = x_i(\tau) Q^\Delta_i.$$ 

Without loss of generality, we can assume that  $x_i(\tau) = [1 ~~ 0]$, i.e. the chain at time $\tau$ is in state $0$. This is because the chain is symmetric, so it does not matter which state we start from. Using this, we get the estimate distribution to be 
\begin{equation}
	\hat{x}_i(\tau+\Delta) = \big[ [Q^\Delta_i]_{00} ~~ 1-[Q^\Delta_i]_{00} \big].
\end{equation}

Further, at time $\tau+\Delta$ the actual state of the chain is $0$ with probability $[Q^\Delta_i]_{00}$ and $1$ with probability $1-[Q^\Delta_i]_{00}$. Thus, the distribution of the actual state at time $\tau+\Delta$ is given by
\begin{equation}
	x_i(\tau+\Delta) = \begin{cases}
		[1~~0], &\text{ with probability }[Q^\Delta_i]_{00}\\
		[0~~1], &\text{ with probability }1-[Q^\Delta_i]_{00}.
	\end{cases}
\end{equation}

Now, suppose that the distance between the actual and estimate distributions is measured using the Kullback-Leibler (KL) divergence. Then,
\begin{equation}
	\begin{aligned}
		\mathbb{E}&\bigg[ D_{KL}\big(x_i(\tau+\Delta) || \hat{x}_i(\tau+\Delta)\big) \bigg] 	\\
		&= [Q^\Delta_i]_{00} D_{KL}\bigg( [1~~0] \bigg|\bigg|  \big[ [Q^\Delta_i]_{00} ~~ 1-[Q^\Delta_i]_{00} \big] \bigg) \\ &+ 
		(1-[Q^\Delta_i]_{00}) D_{KL}\bigg( [0~~1] \bigg|\bigg|  \big[ [Q^\Delta_i]_{00} ~~ 1-[Q^\Delta_i]_{00} \big] \bigg)\\
		&= -[Q^\Delta_i]_{00}\log\big( [Q^\Delta_i]_{00} \big) - (1-[Q^\Delta_i]_{00})\log\big( 1-[Q^\Delta_i]_{00} \big) \\
		&= H\big( [Q^\Delta_i]_{00} \big).
	\end{aligned}
\end{equation} 
Here $H(q) \triangleq -q\log(q) -(1-q)\log(1-q)$ is the binary entropy function.

Now, suppose that the distance between the actual and estimate distributions is measured using the total variation (TV) distance. Then,
\begin{equation}
	\begin{aligned}
		\mathbb{E}&\bigg[ D_{TV}\big(x_i(\tau+\Delta) || \hat{x}_i(\tau+\Delta)\big) \bigg] 	\\
		&= [Q^\Delta_i]_{00} D_{TV}\bigg( [1~~0] \bigg|\bigg|  \big[ [Q^\Delta_i]_{00} ~~ 1-[Q^\Delta_i]_{00} \big] \bigg) \\ &+ 
		(1-[Q^\Delta_i]_{00}) D_{TV}\bigg( [0~~1] \bigg|\bigg|  \big[ [Q^\Delta_i]_{00} ~~ 1-[Q^\Delta_i]_{00} \big] \bigg)\\
		&= [Q^\Delta_i]_{00}\big(1- [Q^\Delta_i]_{00} \big) + (1-[Q^\Delta_i]_{00})[Q^\Delta_i]_{00}  \\
		&= 2[Q^\Delta_i]_{00}\big(1- [Q^\Delta_i]_{00} \big)\\
		&\triangleq g\big( [Q^\Delta_i]_{00} \big).
	\end{aligned}
\end{equation} 
Here $g(x) = 2x(1-x)$. This completes the proof of Theorem \ref{thm:markov_fAoI}.

In addition, we also need to show that the two functions derived above are monotonically increasing. To do so, we will simplify our notation a bit. Let $\mu_0 = [Q^\Delta_i]_{00}$ by $\mu_0$ and $\mu_1 = 1 - \mu_0 = [Q^\Delta_i]_{01}$. Further, let $\nu_0 = [Q^{\Delta+1}_i]_{00} = \mu_0(1-q_i) + (1-\mu_0)q_i$ and $\nu_1 = [Q^{\Delta+1}_i]_{01} = 1 - \nu_0 = \mu_1(1-q_i) + (1-\mu_1)q_i$. We will split the proof into two cases.

\textbf{Case 1 (KL Divergence):} 
Note that the function $x \log(x)$ is convex for all $x > 0$, since $\frac{d^2}{dx^2} (x \log(x))   = \frac{1}{x} > 0, \forall x > 0.$ Using this fact and the definitions of $\nu_0$ and $\nu_1$, we obtain the following inequalities:
\begin{equation}
	\label{eq:ent1}
	(1-q_i){\mu}_0 \log({\mu}_0) + q_i {\mu}_1 \log({\mu}_1) \geq \nu_0 \log(\nu_0),
\end{equation}
\begin{equation}
	\label{eq:ent2}
	(1-q_i){\mu}_1 \log({\mu}_1) + q_i {\mu}_0 \log({\mu}_0) \geq \nu_1 \log(\nu_1),
\end{equation}
Now, we look at the difference:
\begin{multline}
	H\big( [Q^{\Delta+1}_i]_{00} \big) - H\big( [Q^{\Delta}_i]_{00} \big) = \\
	\bigg(
	(1-q_i){\mu}_0 \log({\mu}_0) + q_i {\mu}_1 \log({\mu}_1) - \nu_0 \log(\nu_0)  \bigg) \\ + \bigg(
	(1-q_i){\mu}_1 \log({\mu}_1) + q_i {\mu}_0 \log({\mu}_0) - \nu_1 \log(\nu_1)  \bigg)\\
	\ge 0.
\end{multline}
The inequality above follows by applying \eqref{eq:ent1} and \eqref{eq:ent2}. This proves that the monitoring error grows monotonically with the AoI for KL divergence.

\textbf{Case 2 (TV distance):} Note that the function $2x(1-x)$ is concave for all $x$, since $\frac{d^2}{dx^2} (2x(1-x))   = -2 < 0, \forall x.$ Using this fact and the definitions of $\nu_0$ and $\nu_1$, we obtain the following inequalities:
\begin{equation}
	\label{eq:tv1}
	(1-q_i)2{\mu}_0 (1-{\mu}_0) + q_i 2{\mu}_1 (1-{\mu}_1) \leq 2\nu_0 (1-\nu_0),
\end{equation}
\begin{equation}
	\label{eq:tv2}
	(1-q_i)2{\mu}_1 (1-{\mu}_1) + q_i 2{\mu}_0 (1-{\mu}_0) \leq 2\nu_1 (1-\nu_1),
\end{equation}
Now, we look at the difference:
\begin{multline}
	g\big( [Q^{\Delta+1}_i]_{00} \big) - g\big( [Q^{\Delta}_i]_{00} \big) = \\
	\bigg(
	2\nu_0 (1-\nu_0) - (1-q_i)2{\mu}_0 (1-{\mu}_0) - q_i 2{\mu}_1 (1-{\mu}_1)  \bigg) \\ + \bigg(
	2\nu_1 (1-\nu_1) - (1-q_i)2{\mu}_1 (1-{\mu}_1) - q_i 2{\mu}_0 (1-{\mu}_0)  \bigg)\\
	\ge 0.
\end{multline}
The inequality above follows by applying \eqref{eq:tv1} and \eqref{eq:tv2}. This proves that the monitoring error grows monotonically with the AoI for TV distance as well.

\end{document}